\documentclass[preprint,floatfix,aps,pre,amsfonts,amsmath,amssymb,superscriptaddress,showpacs]{revtex4-1}
\usepackage{color}
\usepackage{graphicx}
\usepackage{dcolumn}
\usepackage{multirow}
\usepackage{bm}
\usepackage{textcomp}
\usepackage{overpic,subfigure}
\begin{document}

\title{Feedback Control of Turbulent Shear Flows by Genetic Programming}

\author{Thomas Duriez}
\email{thomas.duriez@gmail.com}
\affiliation{CONICET, Universidad de Buenos Aires, Ciudad Autonoma de Buenos Aires, Argentina}

\author{Vladimir Parezanovi\'c}
\author{Kai von Krbek}
\author{Jean-Paul Bonnet}
\author{Laurent Cordier}
\author{Bernd R. Noack}\altaffiliation[Also at ]{Institut f\"ur Str\"omungsmechanik, Technische Universit\"at Braunschweig, 38108 Braunschweig, Germany}
\affiliation{ Institut PPRIME, CNRS -- Universit\'e de Poitiers -- ENSMA, UPR 3346, Poitiers, France}

\author{Marc Segond}
\author{Markus Abel}
\affiliation{ Ambrosys GmbH, Potsdam, Germany}

\author{Nicolas Gautier}
\author{Jean-Luc Aider}
\affiliation{ Laboratoire PMMH, CNRS UMR 7636-- ESPCI-ParisTech -- Universit\'e Marie Curie -- Universit\a'e Paris-Diderot, Paris, France}

\author{Cedric Raibaudo}
\author{Christophe Cuvier}
\author{Michel Stanislas}
\affiliation{ Laboratoire de M\'ecanique de Lille, UMR CNRS 8107 -- \'Ecole Centrale de Lille, Villeneuve d'Ascq, France}

\author{Antoine Debien}
\author{Nicolas Mazellier}
\author{Azeddine Kourta}
\affiliation{ Laboratoire PRISME, Université d'Orléans, Orléans, France}

\author{Steven L. Brunton}
\affiliation{ University of Washington, Seattle, USA}

\begin{abstract}
Turbulent shear flows have triggered fundamental research in nonlinear dynamics,
like  transition scenarios, pattern formation and dynamical modeling.
In particular, the control of nonlinear dynamics is subject of research since decades.
In this publication, actuated turbulent shear flows
serve as test-bed for a nonlinear feedback control strategy
which can optimize an arbitrary cost function
in an automatic self-learning manner.
This is facilitated by genetic programming
providing an analytically treatable control law.
Unlike control based on PID laws or neural networks,
no structure of the control law needs to be specified in advance.
The strategy is first applied to low-dimensional dynamical systems
featuring aspects of turbulence
and for which linear control methods fail.
This includes stabilizing an unstable fixed point
of a nonlinearly coupled oscillator model
and maximizing mixing, i.e.\ the Lyapunov exponent, for forced Lorenz equations.
For the first time,
we demonstrate the applicability of genetic programming control
to four shear flow experiments
with strong nonlinearities and intrinsically noisy measurements.
These experiments comprise
mixing enhancement in a turbulent shear layer,
the reduction of the recirculation zone behind a backward facing step,
and the optimized reattachment of separating boundary layers.
Genetic programming control has outperformed
tested optimized state-of-the-art control
and has even found novel actuation mechanisms. 
\end{abstract}


\maketitle

\section{Introduction}

The control of turbulence is of fundamental importance 
for understanding of animal motion and 
for many engineering applications. 
It serves as one of the most complex paradigms 
for methods of the control of dynamical systems in general. 
In fluid dynamics, many engineering applications benefit from feedback control, 
like drag reduction of transport vehicles
or green energy harvesting of wind and water 
flows. In these fields, even small gain in energy savings are very important 
due to the huge 
number of vehicles, turbines, etc. involved. 
Apart from fluids, 
a generic control strategy for complex nonlinear systems
is desired in life sciences, 
medicine, and a vast number of engineering applications.

Apart from the importance of turbulence for applications, it serves as a 
paradigm for the study 
of complex dynamical systems. 
Fluid dynamics have led to many important discoveries and 
poses challenges to physics theory. 
Pattern formation, chaos and bifurcations, would not
have been found and studied without the driving force of turbulent flows.
So, we follow this line of thought and use turbulent flows as prototypic example
for a complex dynamical system. 
Yet, control strategies are best developed for simple nonlinear dynamical 
systems. 
Consequently, we study in this publication
the control of two systems with nonlinear features of turbulence:
(1) two coupled forced oscillators which are linearly uncontrollable, 
and (2) a forced Lorenz system,
where we optimize for mixing, i.e.\ aim at large Lyapunov exponents. 
Based on these results,
we apply our control algorithm to experiments with turbulence. 

For many laminar flows, linear control theory is used 
for the stabilization of a steady solution
by a local linearization of the Navier-Stokes equation.
Corresponding numerical and experimental stabilization studies include
virtually any configuration,
e.g. wakes ~\citep{Roussopoulos1993jfm},
cavity flows~\citep{Rowley2006arfm,sipp2007jfm,illingworth2012feedback},
flows of backward-facing step~\citep{Herve2012jfm},
boundary-layer flows~\citep{Bagheri2009jfm}.

Turbulent flows, however are next to impossible to stabilize, 
because of the huge number of degrees of freedom, 
the non-normal amplification mechanisms, 
complex bifurcations and nonlinear frequency cross talk.
In addition, real-world constraints complicate control, e.g. the cost of energy 
of actuators,
or their blank inability to control to a steady solution, because actuators 
cannot provide enough
energy or do not have the authority to force the fluid into a steady flow.
Linear control fails to handle frequency cross-talk between the coherent 
structures,
the mean flow and the stochastic small-scale fluctuations.
Yet, frequency cross-talk is an important actuation opportunity
as demonstrated by successful wake stabilization with
high-frequency 
actuation~\citep{Glezer2005aiaaj,thiria2006wake,Luchtenburg2009jfm}
and low-frequency forcing~\citep{Pastoor2008jfm}.
For real-world applications, model-based control of an experiment
requires a robust control-oriented reduced-order model
which is still a large challenge at this moment.
Such a reduced-order model needs --- at minimum ---
to resolve  the uncontrolled and controlled turbulent coherent structures
including the transients between them.

Intriguingly, animals have found feedback flow control solutions
neither knowing the Navier-Stokes equation nor reduced models thereof.
Examples are eagles whose feathers near the trailing edge move up to prevent 
stall.
Or dolphins whose compliant skins delay transition and thus decrease drag.
More than 50 years ago, Rechenberg \cite{Rechenberg1973book}
and Schwefel \cite{Schwefel1965diplom} 
have imitated nature's evolutionary principles
to solve engineering flow problems.
Subsequently, evolutionary algorithms have rapidly evolved
culminating in the genetic algorithm \cite{Holland1975book}
and genetic programming \cite{Koza1992}
and powerful methods of machine learning.
In the pioneering wall turbulence simulation of Lee et al. \cite{Lee1997},
a feedback law for skin friction reduction 
was learned in model-free manner employing a neural network.
The current study continues these works
and employs genetic programming as powerful regression technique
for the optimization of the multiple-input multiple-output control laws.

\begin{figure}
\begin{center}
\includegraphics[width=0.5\textwidth,natwidth=610,natheight=692]
{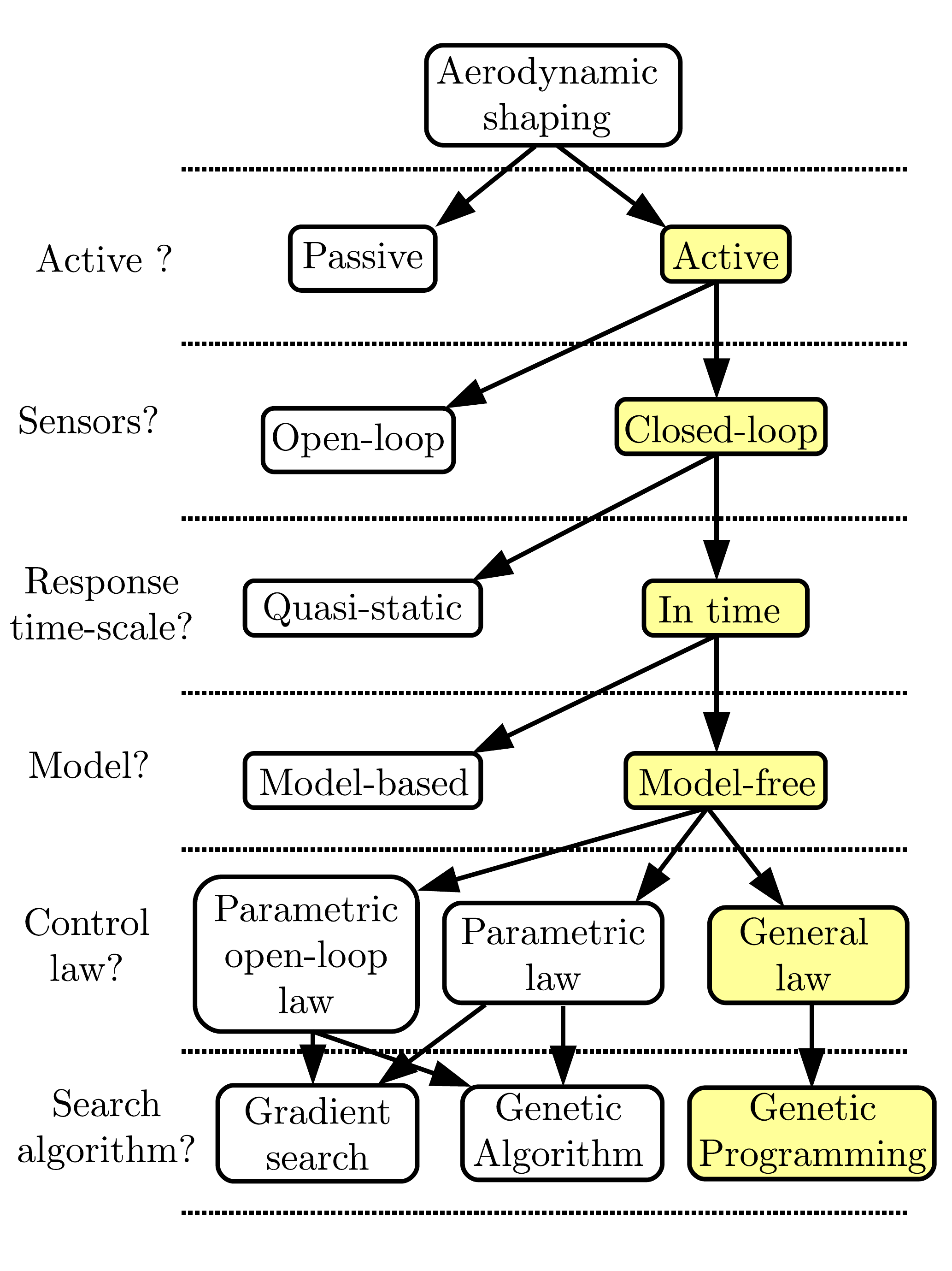}
\end{center}
\caption{Decision tree for control design.}\label{fig:control_decision}
\end{figure}

In the following, we discuss in detail the different approaches on the example 
of turbulence. The paper is organized as follows:
In section~\ref{sec:GP}, we discuss the methodological approach; 
we present our novel model-free approach based on GP for 
controlling complex and turbulent flows.  We motivate the 
use of GP and describe the algorithm, called Genetic Programming Control (GPC). 
Then in section~\ref{sec:proof} we show the possibilities of GP-based 
model-free 
control by using it to stabilize a non-linear system exhibiting frequency 
cross-talk between two oscillators and to enhance the chaos in a periodic 
Lorenz 
system. In section~\ref{sec:exp} we display the successes obtained on 
experimental demonstrators including the mixing enhancement of a turbulent 
shear-layer, recirculation reduction on a backward facing step using real-time 
PIV and separation control on a separating turbulent boundary layer. 
Finally, we  discuss the advantages and drawbacks of the methods and conclude 
in 
section~\ref{sec:conclusion}.

\section{Method}\label{sec:GP}

In this section, we review  control approaches to 
turbulent flows with reference to applications in that field (section \ref{sec:TuCoReview}). 
Then, we describe the control algorithm we chose, where genetic programming 
serves to identify the control law (sections \ref{sec:GP:Algorithm}--\ref{sec:GP:Parameters}).

\subsection{Approaches to control turbulent flows}
\label{sec:TuCoReview}

\subsubsection{Passive or active control}
Many systems have to be modified to improve their aerodynamics performance for 
off-design situations. These modifications are to be achieved as to not disrupt 
the main design objectives of the system. To this end, the field of flow 
control developed many strategies to control 
flows~\cite{Gadelhak1998book,amitay1998aerodynamic,amitay2001aerodynamic,
cattafesta2003review,colonius2001overview}. The nature of the system as well as 
the objective of the control can guide the designer's choices while going 
through a decision tree as depicted in Fig.~\ref{fig:control_decision}. 
Passive control is on many aspects (design and maintenance cost, failure odds 
e.g.) the easiest way to achieve flow control. It consists in adding small 
passive devices or slightly change the geometry of the system (e.g. deflectors, 
vortex generators) in order to reach better flow 
configurations~\cite{cattafesta2011actuators}. The first drawback of passive 
control is that it impacts the system under all operating conditions while the 
control is only required in particular circumstances. For instance the design 
of wings is strongly impacted by take-off and landing conditions, which may lead 
to designs that are sub-optimal for cruising flight conditions. Active control, 
on the other hand, brings as a first advantage the ability to turn the control 
off or on. Then, active control can be characterized by its energy and 
frequency content. Injecting energy improves control authority and makes it 
possible to reach states which are unreachable with passive control. Finally, 
the possibility to add a frequency content allows to act on specific flow 
mechanisms of the flow (usually through instabilities) which leads to higher 
control efficiency.

\subsubsection{Open- or closed-loop control}
Open-loop control has been successful in many applications. Typical instances 
are periodic forcings based on flows' most sensitive 
frequencies.~\cite{chun1996control,thiria2006wake}. Though, such a design can 
only be effective using information that is available during the design phase of 
the control. Should the flow conditions change, then the control can turn 
totally ineffective or can hurt the system performances. A way to solve this 
difficulty is to provide information from sensors to the controller leading to a 
closed-loop control 
design~\cite{taylor2004towards,glauser2004feedback,Rowley2006arfm,
pinier2007proportional,Pastoor2008jfm}. The sensors can be used to evaluate 
global quantities characteristic of the flow e.g. Reynolds number, temperature, 
static pressure, noise level or more specific values e.g. instantaneous 
velocities, dynamical pressure, perturbations. The feedback loop can be employed 
both to evaluate the flow state and the control efficiency which can in turn be 
used to refine the control design. Thus, closed-loop control can exhibit an 
intrinsic robustness when compared to open-loop 
control~\cite{Rowley2006arfm,Pastoor2008jfm}. Furthermore the access to 
dynamical information on the system allows to target the trajectories of the 
dynamical system. 

\subsubsection{Adaptive or in-time closed loop control}

Adaptive control builds on slow adaptation of parameters of a working open-loop 
control, like amplitude or frequency of periodic forcing. It can be implemented 
by exploiting statistical information of the system to evaluate a deterioration 
in the control performance and to correct the control parameters accordingly. It 
has been successfully implemented on many 
applications~\cite{beaudoin2006drag,king2006adaptive,becker2007adaptive}, 
exhibiting remarkable gains in robustness against changes in the flow 
conditions. Adaptive control can be schematized by a succession of effective 
open-loop controls which are selected according to the monitored flow 
conditions. While it has the advantage to allow a comparatively simple technical 
implementation thanks to modest needs in the information flow frequency (which 
is usually one to several order of magnitude lower than the frequency content of 
the relevant physics), its absolute performance can never be better than the 
selected open-loop controls for the considered flow conditions. By allowing an 
in-time information flow to the controller, the system trajectory can be 
inferred in real-time, opening the way to stabilization (e.g. opposition 
control~\cite{choi1994active}, time-delay synchronization~\cite{Pyragas1992}). 
This class contains the  pioneering simulations 
on skin friction reduction of wall turbulence
--- either by stabilizing the whole flow in time~\cite{Moin1994amr}	
or by preventing streaks in the laminar wall 
region~\cite{choi1994active,lee1998suboptimal,Kim2003pf}.

\subsubsection{Model-based or model-free}
\begin{figure*}
\begin{center}
\includegraphics[width=0.85\textwidth,natwidth=610,natheight=620]
{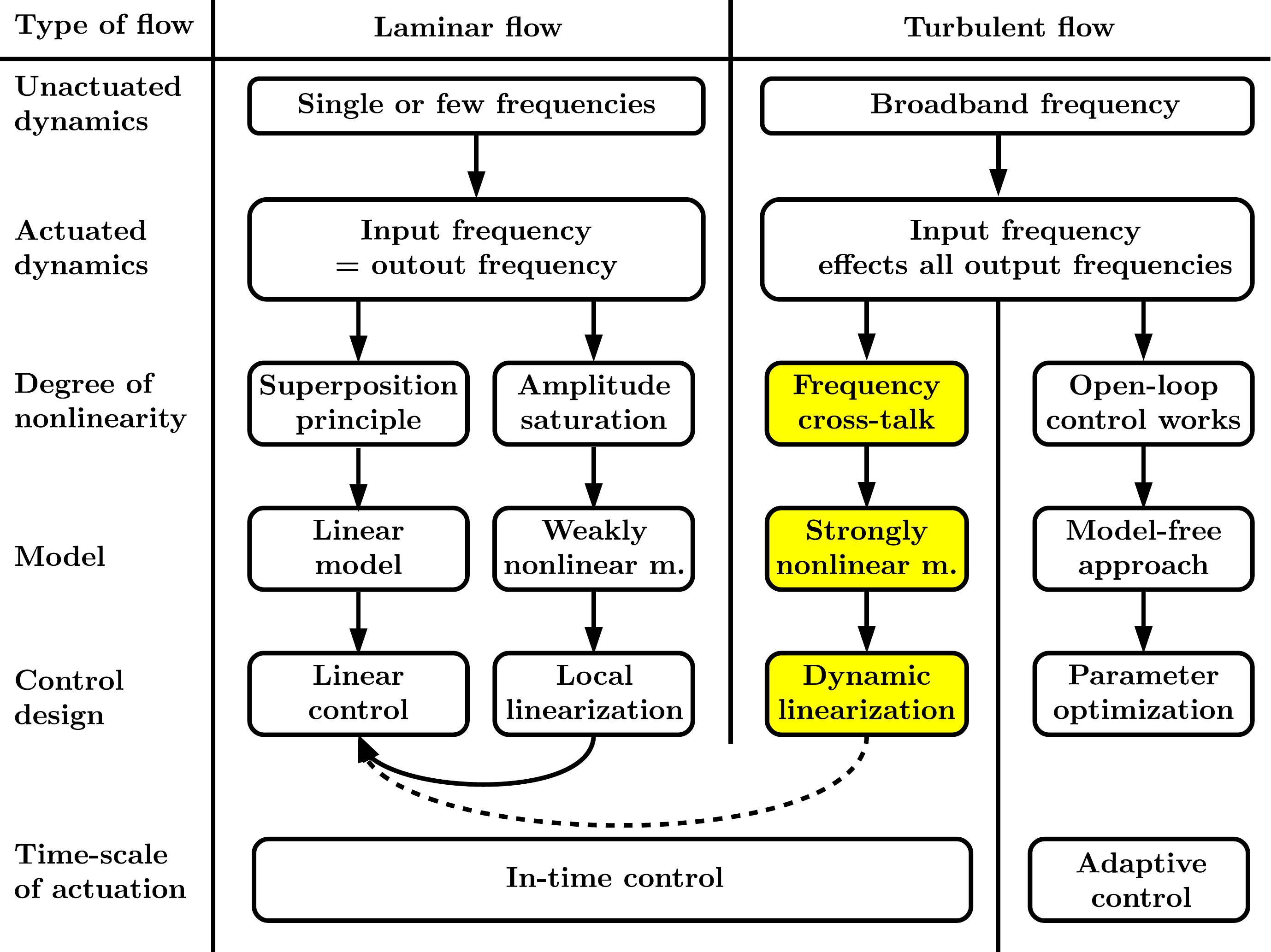}
\end{center}
\caption[Feedback Control Physics]{Physics of closed-loop turbulence control 
strategies with considered key challenges  marked in yellow.
For details see text.
}
\label{Fig:ControlPhysics}
\end{figure*}

Monitoring trajectories and acting accordingly based on incomplete sensor 
information is a challenging task. A first approach is to use a model of the 
flow to both build an observer that can estimate the state of the flow based on 
the information of the sensors, and compute the best possible actuation that 
leads the modeled flow in the desired state. As most flows correspond to 
high-order dynamical systems inadequate for real-time control, a 
control-oriented reduced-order model (ROM) is useful as test-bed for the 
understanding of the actuation mechanism and as enabler for online-computability 
in an experiment. The tools developed in the linear control theory have 
demonstrated numerous successes on flows that are well adapted to model 
reduction~\cite{noack2003hierarchy,samimy2004exploring,rowley2004model,
rowley2006linear}. With increasing flow complexity, this task becomes 
exponentially difficult if at all possible. These challenges are summarized in 
Fig.~\ref{Fig:ControlPhysics} and detailed below.
The next step would be to asses if a model of the flow is needed or at all 
available. The sparse experimental literature  on in-time turbulence control can 
be explained with the immense modeling and control challenges.

Many laminar shear flows generally exhibit one or few frequencies.
The actuation frequency (input)
is recorded by the sensors (output).
Steady solutions at low Reynolds numbers have an actuation dynamics
which may approximately satisfy a \textit{superposition principle:}
The sum of two actuations yields the sum
of the individual responses in the sensors.
In this case, a linear model approximates
the flow behavior
and can be used for linear control design.
At larger Reynolds number, flow mechanisms like
periodic vortex shedding have evolved
due to an instability and it are limited by a nonlinear amplitude saturation 
mechanism.
In this case, the superposition principle is no longer valid,
but control may be based on a locally linearized model parameterized by the base 
flow
\cite{Noack2011book,Holmes2012book}.

Continuing with the right side of Fig.~\ref{Fig:ControlPhysics},
turbulent flows exhibit  a broadband frequency dynamics seen by the sensors
already in the natural case.
A monochromatic forcing changes the whole spectrum
due to frequency cross-talk,  like vortex pairing (inverse cascade)
or the normal cascade to smaller structures.
It is a non-trivial challenge to frame this actuation response in a 
reduced-order model.
In many studies,
a periodic forcing is optimized
in a model-free adaptive manner
(see right-most column of Fig.~\ref{Fig:ControlPhysics}).
In some cases, the nonlinear frequency cross-talk
can be modeled and exploited for a 'dynamic linearization'
leading to a linear control law.
One example is a turbulent flow dominated by a natural shedding frequency and 
mitigated by a different actuation frequency~\cite{Glezer2005aiaaj}.
In this case, the flow
can be described by generalized mean-field model~\cite{Luchtenburg2009jfm}.
For more complex dynamics,
the search for a model-based control strategy constitutes a large challenge. An 
attempt at obtaining a linear model for the mixing layer flow presented in 
section~\ref{sec:TUC} is described in Appendix.
The sparse experimental literature  on in-time turbulence control can be 
explained with these immense modeling and control challenges.
	
On the other hand, a model-free approach can be investigated where the control 
law has to be determined based on realizations of the process.

\subsubsection{Choice of a control law}
Following Fig.~\ref{fig:control_decision}, the next decision would be on 
whether the control law should be found by parameter identification of an 
open-loop forcing, e.g. periodic forcing, or by parameter tuning of an in-time 
control law, e.g. PID feedback, or by structure identification of a general law. 
Parametrization of open-loop forcing could correspond e.g. to the determination 
based on the flow condition of the optimal frequency, amplitude and duty cycle 
of an harmonic forcing. A parametric law could correspond to the determination 
of the coefficients of the Taylor expansion of a control law, or the weights of 
the connections in a neural network. Determining a general law is the ultimate 
grail as this determination is likely to reveal the enabling, possibly 
nonlinear mechanism which allows an effective control. This is the approach 
considered in this contribution.

The search algorithm used to obtain the control law is crucial. A gradient 
search method is able to find local minima around the initial operating 
conditions. Examples are extremum and slope seeking as well as auto-tuning PID 
controllers. The known drawback of gradient-based methods is their high 
sensitivity to local minima. Depending on the topology of the cost function 
graph, this can lead to the inability to find a sufficiently efficient control 
or to hysteresis under changing conditions. On the other hand, evolutionary 
algorithms like genetic algorithms (GA) and genetic programming (GP) have both 
exploration and exploitation mechanisms. GA and GP which belong to the machine 
learning field can be used for solving global optimization problems, giving the 
possibility to find a global extremum even when a local minimum is first 
detected. It is to be noted that Monte Carlo methods also treat successfully the 
local minima problem but those methods lack an effective exploitation phase 
which leads to prohibitive costs for a large search space. Machine learning 
methods have been used successfully to derive micro-controllers for robotics, 
computer programs, optimal shapes, image recognition and data-based decision to 
name a few examples~\cite{Wahde2008}. The evolutionary algorithms from machine 
learning have the advantage of being most performant when the search space is 
complex, especially when numerous local extrema are present. As GA can only 
optimize parameters, it cannot be used to derive the expression of a general 
function, but can be employed as a search algorithm in a neural network. GP is 
able to construct general functions from a set of given user-defined functions. 
This algorithm is the best candidate to determine in-time closed-loop control 
laws in a model-free framework.

\subsection{ Genetic Programming Algorithm}
\label{sec:GP:Algorithm}

This algorithm relies on GP to derive feedback control laws for any complex 
dynamical system. In the sequel this algorithm is called Genetic Programming 
Control (GPC). We first formulate the control problem in a dynamical system's 
framework. After motivating the use of GP with respect to other popular 
evolutionary algorithms, we present the different steps used in the GP process 
to determine an effective control law. Finally we discuss the choice of 
parameters as well as the stopping criteria used to judge the convergence.

In the following, we restrict the description to ordinary differential equations 
for reasons of comprehensibility. The generic multiple-input/multiple-output 
(MIMO) system is represented in phase space by the vector $\mathbf{a}\in 
\mathbb{R}^{n_a}$, the state is monitored by sensors $\mathbf{s}\in 
\mathbb{R}^{n_s}$, and controlled  by actuators $\mathbf{b}\in 
\mathbb{R}^{n_b}$,
\begin{equation}
\frac{d\mathbf{a}}{dt} = \mathbf{F} \left( \mathbf{a}, \mathbf{b}\right)\, ,\;
\mathbf{s}       = \mathbf{H} \left( \mathbf{a}\right)\, ,\;
\mathbf{b}  = \mathbf{K} \left(\mathbf{s}\right) \,,
\end{equation}
with $\mathbf{F}$ denoting a general nonlinear function, $\mathbf{H}$ the 
measurement function, and $\mathbf{K}$ the sensor-based control law. This law 
minimizes the state- and actuation-dependent cost function:
\begin{equation}
J = J(\mathbf{a},\mathbf{b}).
\end{equation}
The cost function value grades the performance of a control law $\mathbf{K}(\mathbf{s})$. 
The lower the value of the cost 
function, the better the control law solves the problem.
\begin{figure*}
\begin{center}
\includegraphics[width=0.8\textwidth,natwidth=610,natheight=692]
{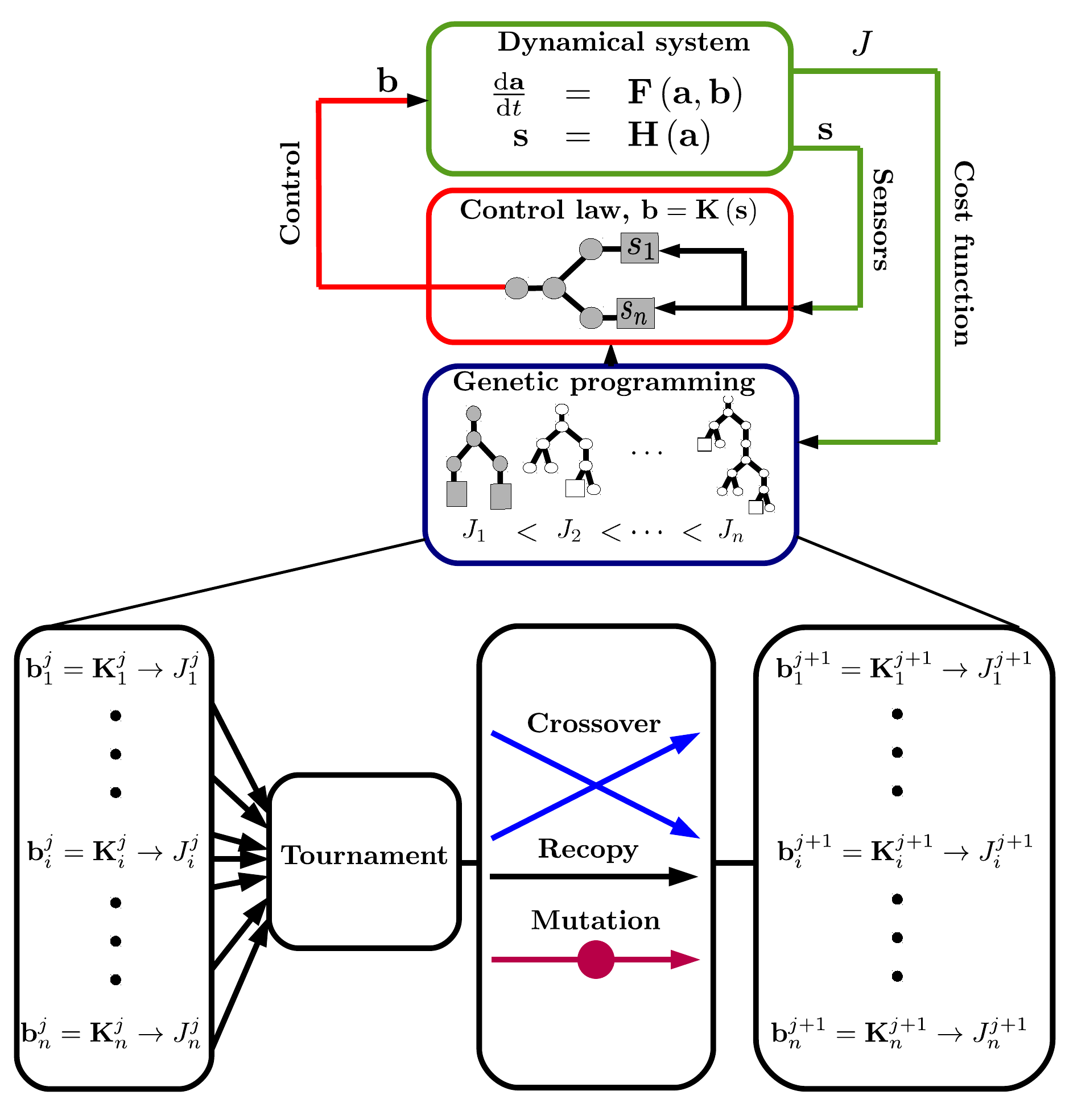}
\end{center}
\caption{Model-free control design using GP. During the learning phase, each 
control law candidate is evaluated by the dynamical system or experimental 
plant. This process is iterated over many generations of individuals. At 
convergence, the best individual with the smallest cost function value (in gray 
in the GP box) is used for control.
} 
\label{fig:GPprocess}
\end{figure*}

\subsection{Genetic Programming}
\label{sec:GP:GP}
For determining the expressions of the optimal feedback laws, potential 
algorithms naturally arise from machine learning~\cite{murphy}.  
Machine learning  algorithms can be roughly categorized as classification, 
regression, clustering and dimensionality reduction. Three algorithms are 
especially suited to learn a system behavior from data: Genetic Algorithms (GA), 
Artificial Neural Network (ANN) and Genetic Programming (GP). 
GA~\cite{milano2002clustering} and ANN~\cite{Lee1997} have both been used in a 
control context, though, without structure identification of the control 
law. A GA can only guess parameter values, and as such, can only be used 
assuming the structure of the control, which is obtained by other means
before optimization. An ANN is, strictly speaking, the 
arrangement of perceptrons, their interaction and their contained sigmoid 
functions that define an input/output system. The algorithm used to learn the 
weights of the links between perceptrons can be freely chosen among a pool of 
methods. If the algorithm is not evolutionary (e.g. reinforcement by error 
back-propagation in the hidden layers, the most common algorithm used), then the 
learning algorithm is similar to a gradient-based search, which means that the 
process is sensitive to local extrema of the search-space. If the algorithm is 
evolutionary (e.g. GA) then we obtain a process which is able to deal 
successfully with local minima. The main drawback with ANN is 
that it is hard to process the result  analytically.
GP, however, yields a function as a possibly nonlinear combination of 
mathematical operations. Consequently, one can read and analyze the result 
in order to understand how the learned mechanism actually works. 
The evolutionary principle is the same as in genetic algorithm, combining phases 
of search-space exploration and phases of convergence toward an extremum. 
Whereas GP has been extensively used to design computer 
programs~\cite{koza1994genetic}, micro-controllers in 
robotics~\cite{lewis1992genetic,nordin1997line} and to 
identify a dynamical 
system~\cite{watson1996identification,willis1997systems,gray1998nonlinear}, to 
our knowledge, it has never been used before in a model-free control 
strategy for dynamical systems. In the following sections, we detail out 
approach (Fig.~\ref{fig:GPprocess}).

\subsection{First generation}
\label{sec:GP:First}
\begin{figure}
\begin{center}
\includegraphics[width=0.4\textwidth,natwidth=610,natheight=620]
{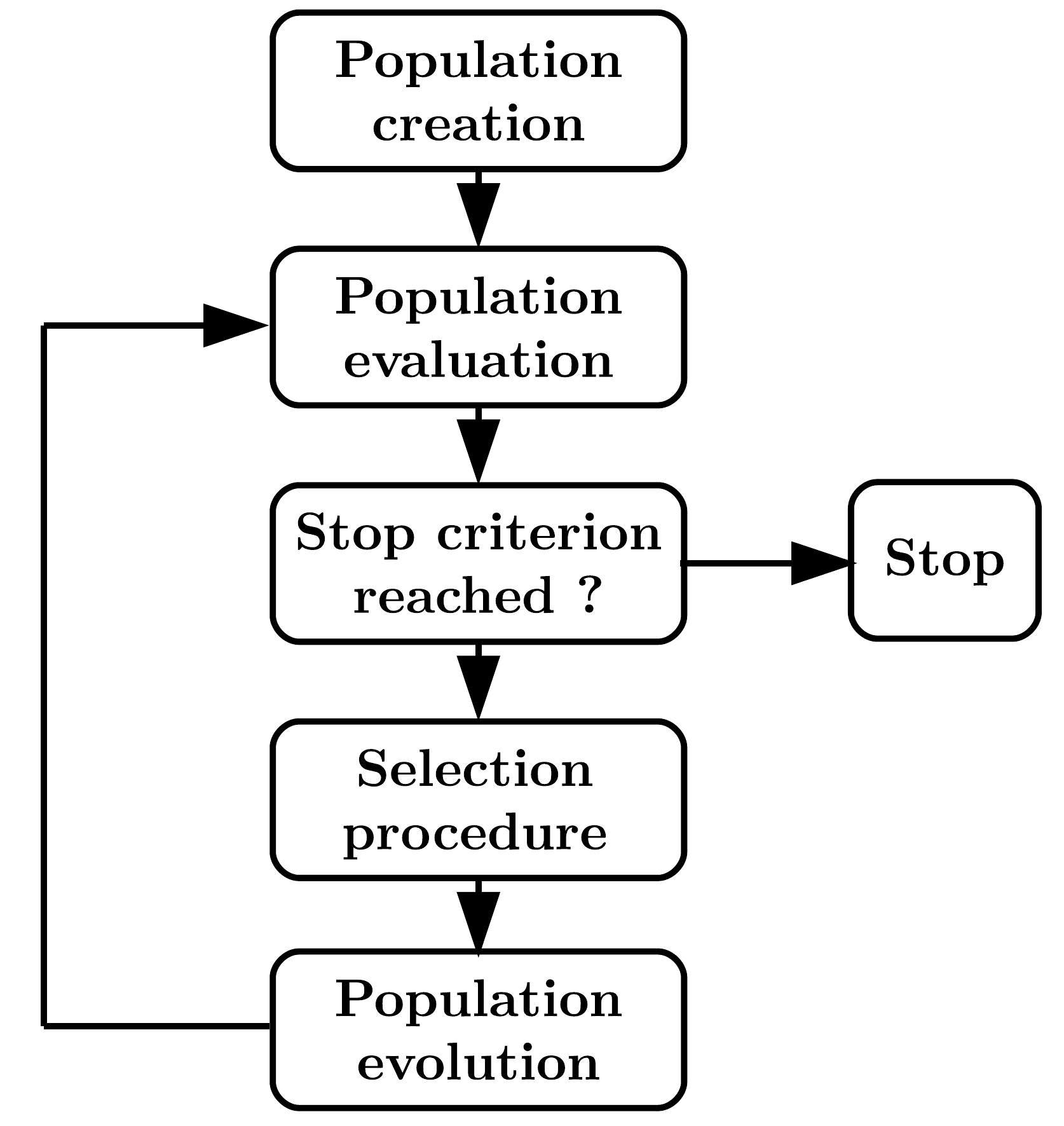}
\end{center}
\caption{Flowchart for the Genetic Programming algorithm.}\label{fig:flowchart}
\end{figure}
The flowchart for GP is given in Fig.~\ref{fig:flowchart}: a set (called 
population) of $n$ control laws (called individuals) is evolved by genetic 
operations: replication, crossover and mutation. Selection is achieved with 
respect to how the individuals behave in terms of minimization of the cost 
function $J$. In the following we detail the algorithm, beginning with the 
population creation.
Several representations can be used to manipulate functions inside the Genetic 
Programming algorithm (e.g. trees, linear programming). We chose to use a 
tree-like representation. The two main advantages of this representation are to 
be human-understandable at a glance, and to be easily synthesized and 
manipulated with a computer. We note that, due to the recursive structure,  
functional programming languages are particularly well suited, as Haskell or LISP.  

\begin{figure}
\centering
\includegraphics[width=0.35\textwidth,natwidth=610,natheight=620]
{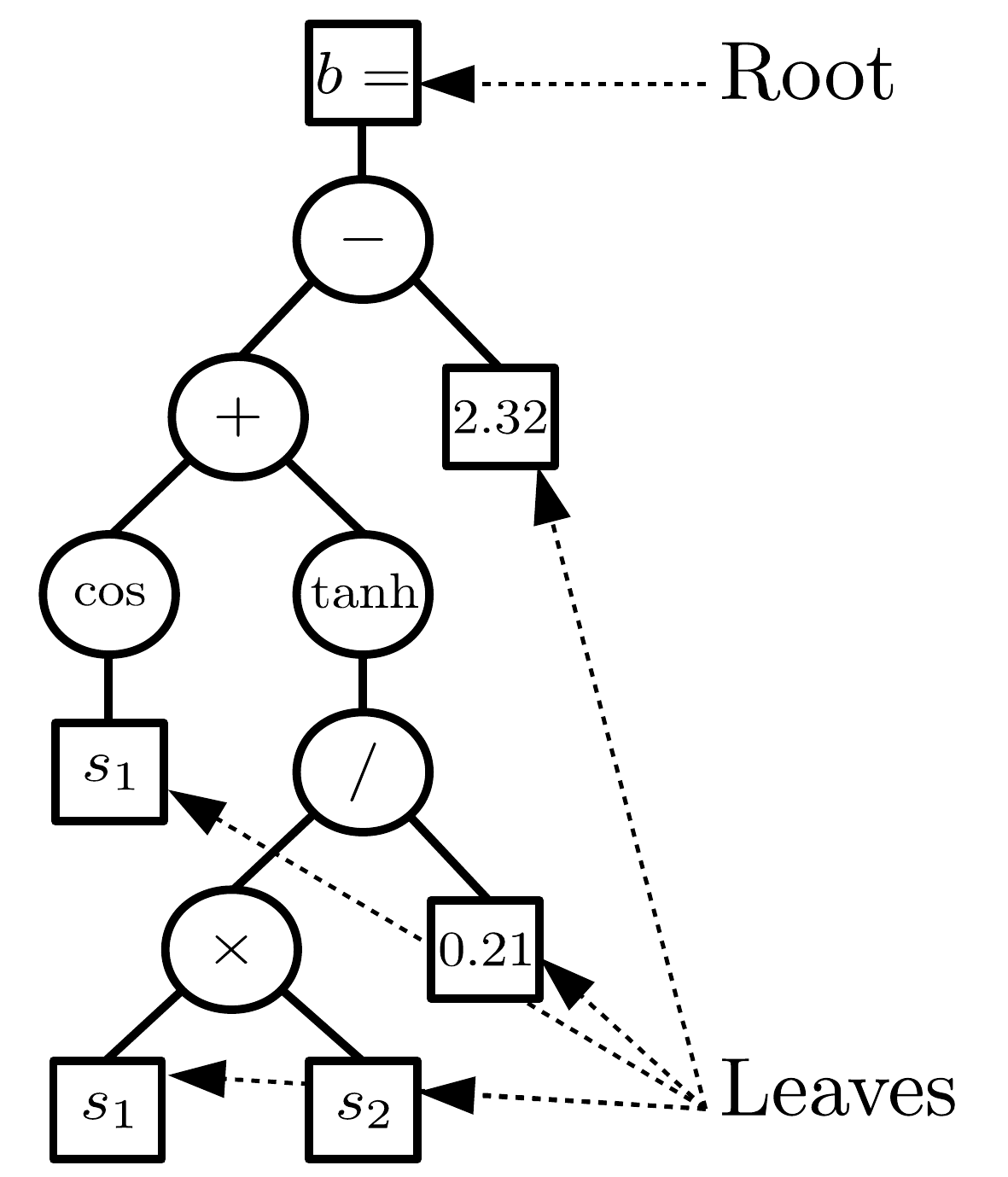}
\caption{An expression tree representing the function $b(s_1,s_2)=\cos(s_1) + 
\tanh((s_1\times s_2)/0.21) - 2.32$.}
\label{fig:exptree}
\end{figure}

In the GPC algorithm, the control laws exist as 
expression-trees (see Fig.~\ref{fig:exptree}). Expression-trees are usually 
built from a number of elementary functions which can take any number of 
arguments but return a single value. Typical examples are 
$(+,-,\times,/,\sin,\tanh,\ldots)$. 
Here, the arguments of these constituents
are sensors values, constants or outputs of sub-trees.
A tree can be built from the final output value (the root of the tree) to 
terminal elements that can take no argument (the leaves). 
A typical visual representation of an expression-tree can be seen in 
Fig.~\ref{fig:exptree}. This tree represents the function:
\begin{equation}
b(s_1,s_2)=\cos(s_1) + \tanh((s_1\times s_2)/0.21) - 2.32,\label{eq:exptree}
\end{equation}
where $s_1$ and $s_2$ represent the time-varying sensor values. A useful way to 
handle an expression tree is to write it recursively, e.g. in Polish notation; 
Equation~\eqref{eq:exptree} is then written as 
\begin{displaymath}
(- \>\> (+ \> (\cos \>\> s_1) \>\> 
(\tanh \>\> (/ \>\> (\times \>\> s_1 \>\> s_2) \>\> 0.21))) \>\> 2.32). 
\end{displaymath}
Though less understandable by the human eye, recursive algorithms can generate, 
derive, manipulate and evaluate these expressions.

The generation process of any individual starts at the root. Then a first 
element is chosen from the pool of admissible elements composed of the basis 
functions and operators, the sensors and the constants. If a basis function or 
operator is chosen,  
new elements are added  as arguments, the process is iterated to include their 
arguments until all branches have leaves. This process is subject to 
limitations. A given tree-depth (the maximum distance between a leaf and the 
root) can be prescribed by preventing the last branch to generate a leaf before 
aforementioned tree-depth and by enforcing the generation of leaves when the 
tree-depth is reached. Similarly it is possible to ensure that each branch 
reaches the same given tree-depth which generates a full density tree. GPC can 
implement any kind of these distributions, from a fully random generation to a 
given tree-depth distribution with given proportion of dense and less dense 
individuals. The first generation starts with a distribution of rather low
  tree-depth (2 to 6) and a $1:1$ distribution of dense and less dense 
individuals. This choice generally ensures enough diversity for the creation of 
next generations. The initially low tree-depth takes into account that the 
genetic operations (see section~\ref{sec:GP:Operations}) have a tendency to make the 
trees grow. 
This phenomenon is known as bloating of the trees.  
Though not used here, we note that it is also possible to 
manipulate the content of the trees by weighting the probability of each 
element. We do not introduce a priori control laws, though in principle possible. 
To enforce diversity, an individual is discarded if it already exists in the 
current population.

\subsection{Evaluation}
\label{sec:GP:Evaluation}
After the first generation of the population is generated, the individuals are 
evaluated according to the value of $J$ which is computed for each individual. 
Usually, the evolution of the dynamical system is evaluated over a sufficiently long time interval
and $J$ is the value of an integral. If the evolution equation of the dynamical 
system is known, then this can be achieved through integration (e.g. 
Navier-Stokes solver). If the system is experimental then the value of $J$ is 
computed by means of recording relevant data throughout one instance of the 
experiment using the individual under consideration.

As imposed by the selection process presented in
section~\ref{sec:GP:Selection}, the only restriction to the design of the
cost function is that a ranking of the values of $J$ assigned to each
individual must be possible. This allows the use of expressions which might
not be differentiable or even continuous. As stated in
section~\ref{sec:GP:Algorithm}, the algorithm assumes that $J$ is positive
and must be minimized. Also for the sake of comparison with other control
methods like Linear Quadratic Regulator, it is convenient to design the
cost function in a similar way.
In the case of several objectives, one can use regularization by adding 
more objectives using a parameter,  the control law contains
appropriate weights:
\begin{equation}
J=J_a+\gamma J_b,
\end{equation}
where $J_a$ is a given measure on the system that need to be minimized in order 
to achieve the control objectives and $J_b$ a value associated with the cost of 
actuation. More objectives can be added, e.g. complexity of the 
function. The coefficient $\gamma$ is 
determined in order to give priority to some objectives. A rule of thumb is to 
set it in order to establish admissible trade-offs between the objectives.

One big problem in the correct evaluation of 
the cost function consists in the integration time needed to gather sufficient statistics
for an accurate evaluation of $J$. This applies in the first place to experimental runs,
which might be expensive. If the recording time is not long enough, the evaluation of one individual 
does not necessarily return the same cost function value for all instances. Of course, that 
holds for huge numerical runs, too.

So, a  large error on the value, either due to a measurement error or an exceptional 
performance of a non robust control law can lead to mistakenly grade a control 
law. As explained below, the search-space is primarily explored around the 
individuals that perform the best. If an intrinsically low-performing individual 
gets a mistakenly good evaluation, then the whole process is endangered. This is 
why a reevaluation process is implemented on the best individuals. Here, the 
five best individuals are evaluated five times, and their cost function is 
averaged. This procedure ensures that the best performing individuals are more carefully 
evaluated to avoid that the search process gets stirred in a wrong direction.

\subsection{Selection}
\label{sec:GP:Selection}

After the evaluation, the population evolution starts: 
each individual is subject to one of the four ``genetic'' operations,
reproduction, crossover, mutation, elitism, explained in detail below in section~\ref{sec:GP:Operations}.
This  process is called tournament.  
Each time an individual needs to be selected for a genetic operation, a number 
$n_p$ of individuals is randomly chosen from the last evaluated 
generation. From this set of individuals, the one with the smallest cost function value is 
selected.
As the population size $n$ is fixed, $n$ selection tournaments are run each time 
a new generation is created. The population is ranked by decreasing cost 
function value. Because typically
$n>>n_p>1$, the probability for individual $i$ to win a tournament is 
$((n-i)/(n-1))^{n_p-1}$. On average, each individual  enters $n_p$ 
tournaments, so if an individual has a rank of $x\times n$ in the generation 
($x\in \left[0\ldots 1\right]$), then its ``genetic content'' contributes 
roughly to $n_p\times (1-x)^{n_p-1}$ new individuals. A typical value for $n_p=7$ 
only the first half of the ranked generation contributes to the next generation. Individuals 
which are ranked higher can still contribute, but these are rare events. In this 
selection procedure, choosing $n_p$ sets the harshness of the selection. On the 
other hand, this selection procedure does not take into account the global 
distribution of the cost function values like in other selection processes such 
as a fitness proportional selection. If an individual performs much better than 
the rest of the population it does not have better selection chances than an 
individual that barely performs better than the rest of the population. This 
choice ensures that the diversity in the population does not disappear too fast 
with an associated trade-off in convergence speed.

\subsection{Genetic operations}
\label{sec:GP:Operations}

\begin{figure}
\begin{tabular}{c}
\includegraphics[width=0.4\textwidth,natwidth=610,natheight=692]{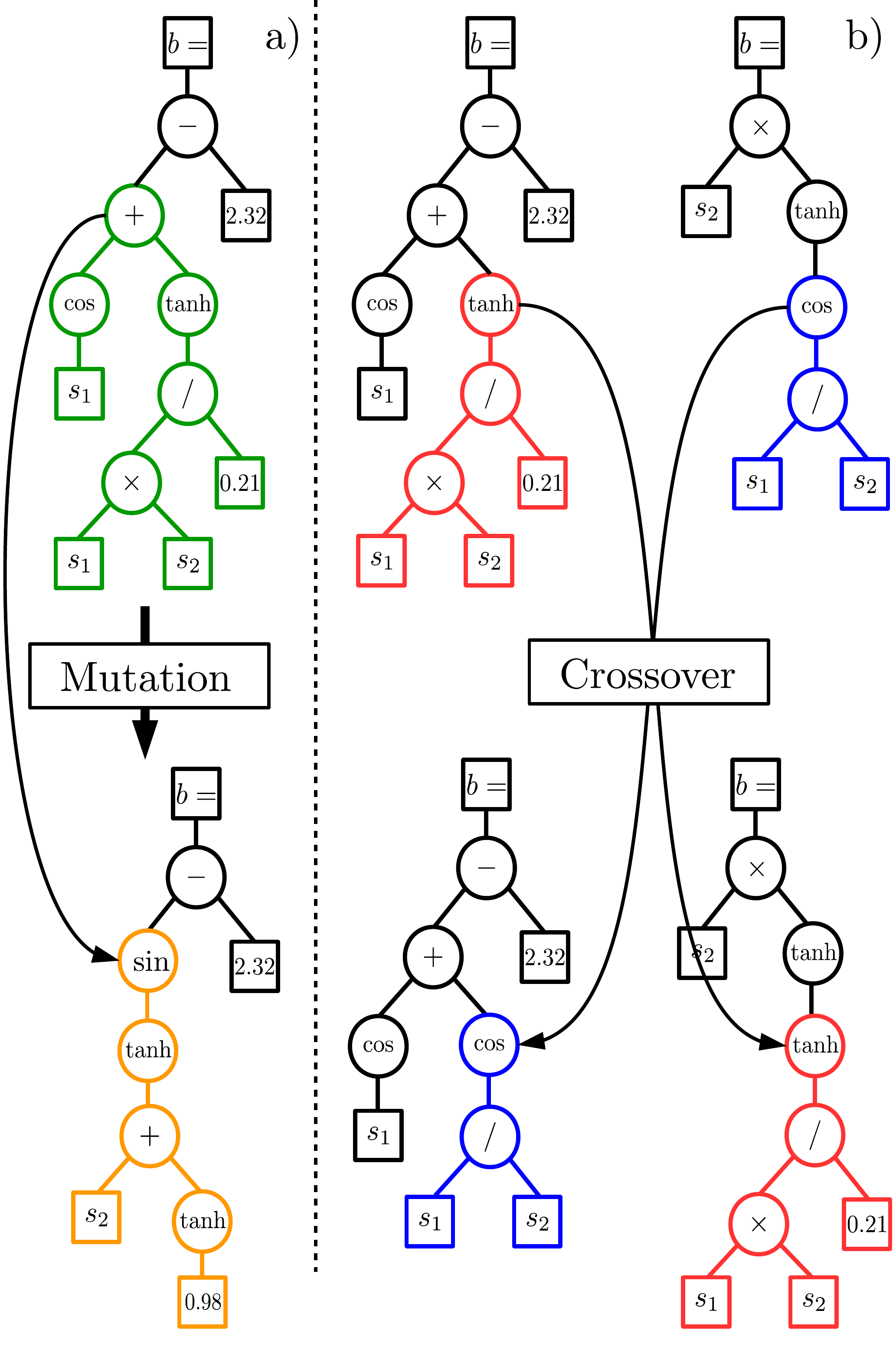}
)\\
\includegraphics[width=0.4\textwidth,natwidth=610,natheight=692]
{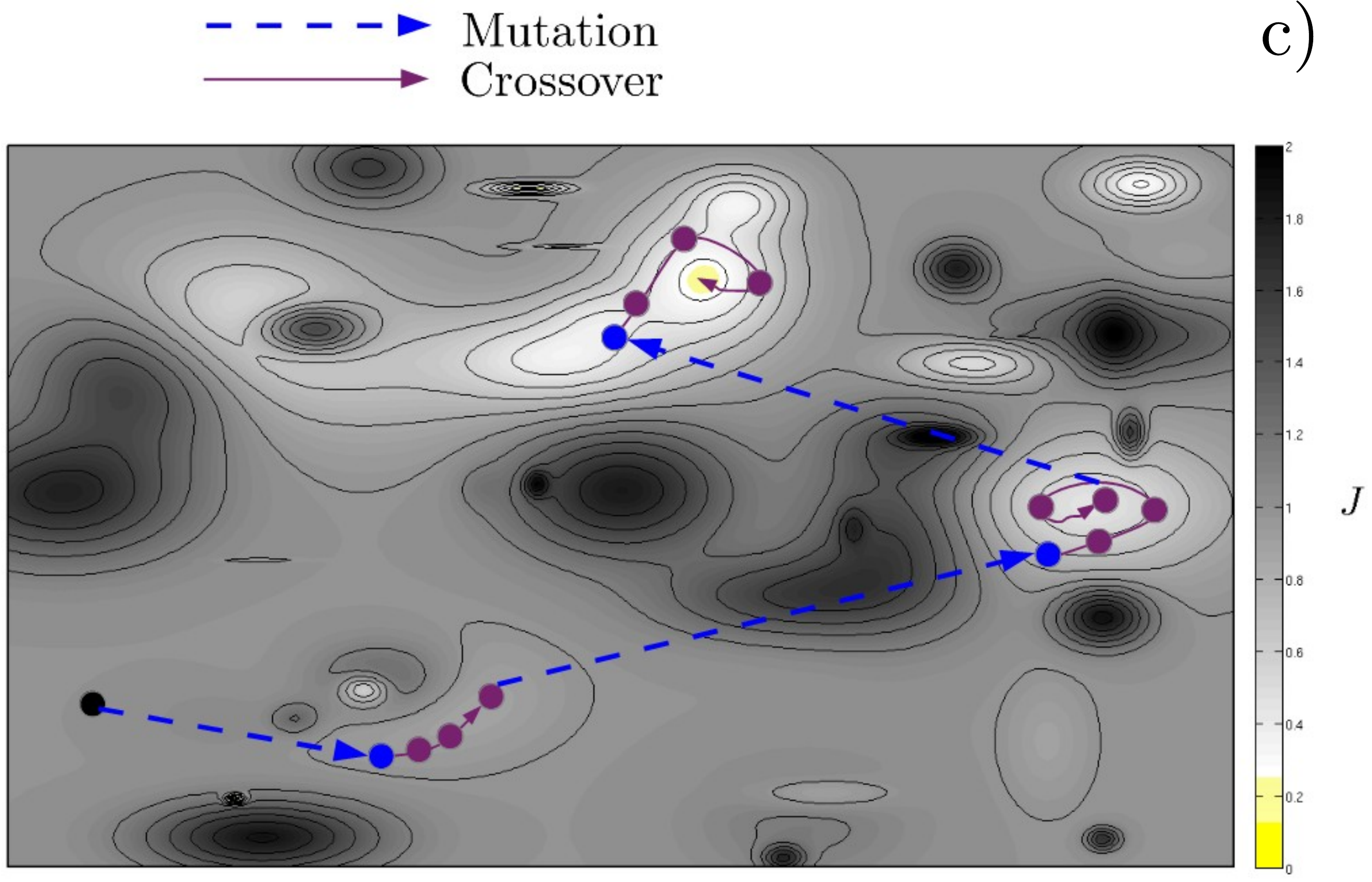}
\end{tabular}
\caption{Top: example of mutation (a) and crossover (b) genetic operations. 
Bottom: Schematic representation of the search-space exploration by mutation 
(dashed lines) and crossover (solid line) by following one branch of the 
genealogy of the best individual (c). While mutations enable large-scale 
exploration, crossovers support the convergence to local 
minima.}\label{fig:crossmut}
\end{figure}

Typically, genetic operations comprise  
 \textit{elitism}, 
 \textit{reproduction}, 
 \textit{mutation} and 
\textit{crossover}.

\textit{Elitism} consists in copying the $n_e$ best individuals of the evaluated 
population to the next generation. This operation ensures that the best control laws stay in the 
population. Once the elitism process is finished, $n-n_e$ individuals are
generated through reproduction, mutation and crossover. The probability of these 
processes is respectively $P_r$, $P_m$ and $P_c$ with $P_r+P_m+P_c=1$. 
\textit{Reproduction} copies the selected individual to the next generation. This is analogue to 
elitism, with lower rank, though.
\textit{Mutation} replaces an arbitrarily chosen subtree by a randomly generated 
new subtree. For that, the same procedure is used as in the creation of the 
first generation. Finally, \textit{crossover} uses two selected individuals and 
exchanges one randomly chosen subtree between them.
 
\textit{Reproduction} ensures a certain stability of the convergence process: it guarantees 
that a part of the population stays in the vicinity of explored local minima of 
the search-space, keeping potentially useful individuals in the population,
exploiting them further before they are discarded. Crossover and mutation are 
responsible respectively for the exploitation and exploration of the 
search-space. As we progress among the generations, the probability to cross 
similar individuals increases: the best individual  propagates its genetic 
content $n_p$ times on average. If this content allows the concerned individuals 
to stay in the first positions of the ranking, it is replicated about 
$n_p^k\times P_c$ times after $k$ generations. Then crossovers of individuals 
selected in the top of the ranking  soon cross similar individuals and 
explore the vicinity of the dominant genetic content. On the other hand, 
mutations introduce new genetic content in the population, hence allowing 
large-scale exploration of the search-space. Mutation and crossover operations 
are displayed in Fig.~\ref{fig:crossmut}. Figure~\ref{fig:crossmut} (bottom) 
illustrates how an evolutionary algorithm explores the search-space for a 
bi-dimensional problem with local minima: the association of these operations 
allows to explore around the local minima found while still exploring the 
search-space for better solutions.

\subsection{GP parameters and stopping criteria}
\label{sec:GP:Parameters}

There are no general rules to choose optimal parameters of 
evolutionary algorithms. 
A common practice is to check the optimality of the solution offered by GP by 
reproducing the process a number of times
using different sets of parameters.
This way, a guiding statistics can be obtained. 
The main impact of the parameters' 
modifications is on the ratio between exploitation (i.e. convergence) and 
exploration of the search-space. Monitoring the evolution of the evaluated 
populations is the best way to fine--tune the GPC process. Now, we discuss the 
role of the GPC's parameters:
\begin{itemize}
\item Population size $n$: the more individuals define the first generation, the 
more the search-space is explored from the intialization. On the other 
hand, a large initial population implies more evaluation time without any 
convergence. 
Let us consider $1000$ evaluations. If only the first generation is evaluated, 
then it is equivalent to a Monte Carlo process. 
Alternatively, one could devote $1000$ evaluations
to $20$ generations with $50$ individuals. 
This implies $20$ convergence steps around the best found individuals.
\begin{figure}
\begin{center}
\includegraphics[width=0.4\textwidth,natwidth=610,natheight=620]{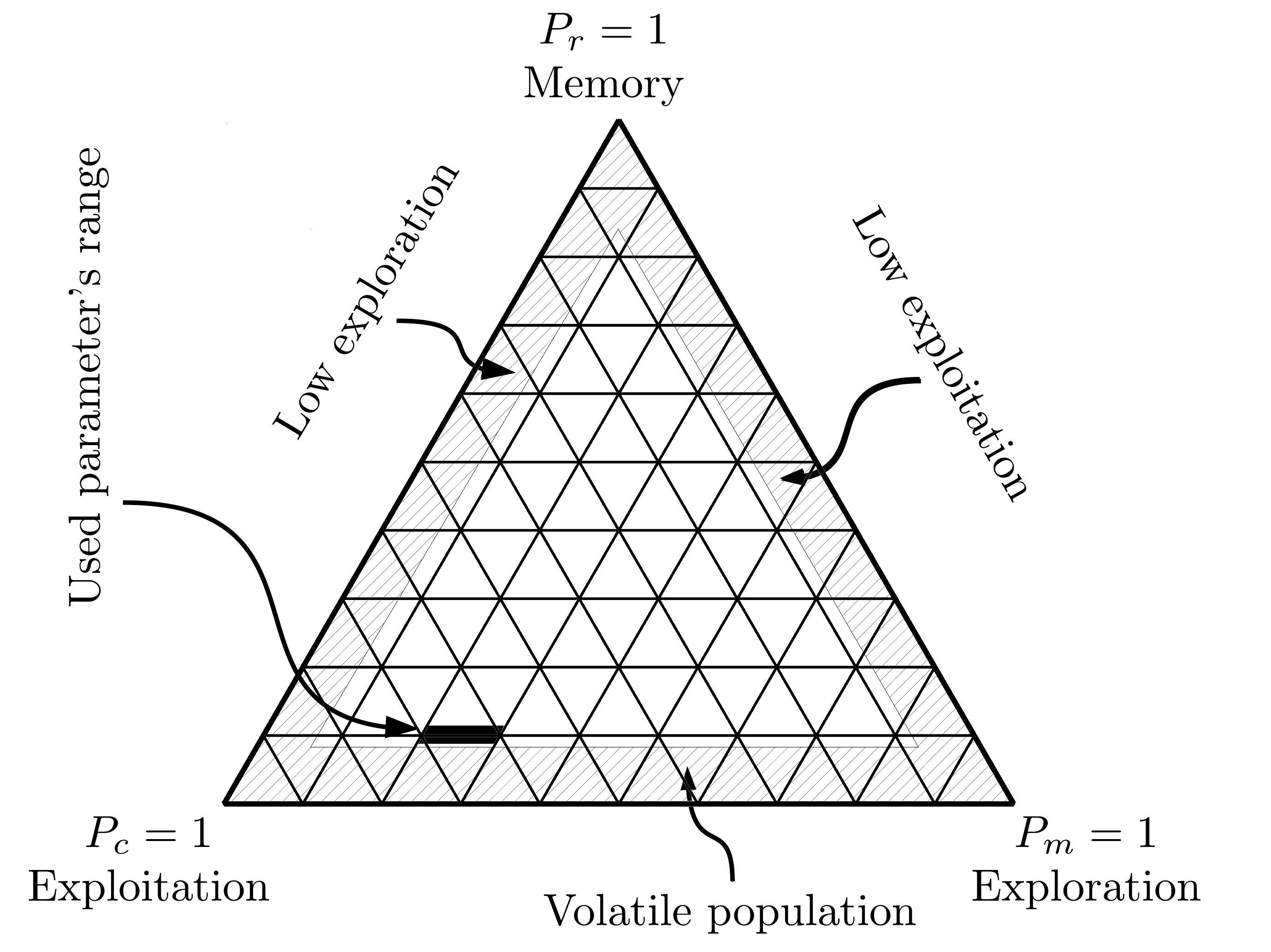}
\end{center}
\caption{Probability selection for the genetic operations. Each line parallel to the opposite side of a summit removes $10\%$ of probability to the genetic operation considered at the summit. There is no globally 
optimal parameter selection. Depending on the problem, diversity or convergence 
needs to be modified. The range of parameters used in the present study is 
represented by the black area. Hatched areas represent the parametric space 
where essential aspects of the evolutionary algorithm are endangered with 
volatile population translating in previous winning options to be forgot once 
a better solution is found (not enough replications), low convergence (not 
enough crossovers) or low exploration (not enough 
mutations).}\label{fig:prob_chart}
\end{figure}
\item Genetic operation probabilities ($n_e$, $P_r$, $P_m$, $P_c$): 
Elitism is encouraged as it ensures that the $n_e$ best performing individuals 
remain in the population. 
Reproduction, mutation and crossover probabilities parametrize the relative 
importance between exploration and exploitation (Fig.~\ref{fig:prob_chart}). 
A large mutation rate $P_m$ implies large-scale exploration and thus population 
diversity. If all individuals in the population share a similar expression and 
have almost similar costs then the mutation rate has to be increased. On the 
other hand, the crossover rate influences the convergence speed. If the 
individuals stay all different and the cost function value histogram does not 
show a peak around the lowest value, then the convergence is not sufficiently 
rapid and the crossover probability needs to be increased. Finally the 
reproduction ensures that a given search-space area is explored during a 
certain number of generations. This parameter ensures as well 
diversity and exploration of different areas.
\item The selection procedure influences also the diversity of 
the following generation. The number of individuals $n_p$ that enter a 
tournament directly influences the number of individuals that contribute to 
the next generation. Reducing $n_p$ increases the diversity while increasing it 
accelerates the convergence.
\item The choice of the elementary functions is intimately linked to the problem 
at stake. This question should be determined in concordance with the choice of 
the sensors and actuators.
\item The maximum number of generations to compute is eventually determined by 
the available testing time. 
A stopping criterion can end the iterations prematurely,
for instance if the optimal solution is reached ($J=0$)
or if the average and minimum of the cost function distribution do not evolve 
anymore.
\end{itemize}
Often, a high rate of crossover and a low rate of mutation is favored to allow for 
a large variety of a initial population as large as possible. However, as stated above,
the concrete choice is problem-dependent. With such choice a fast convergence may be obtained, and
are particularly suited if the evaluations can be parallelized.
In experiments, however, the total time of evaluation is critical and one cannot 
afford a large population. Furthermore the uncertainties in the cost function 
evaluation do not allow to achieve a convergence below the  measurement error.
The best experimental compromise found was to deal with reduced populations (in 
the order of $50$ to $500$ individuals) associated with a high mutation rate 
(from $25$ to $50\%$). Though we decided to keep these values constant during 
the course of each experiment, further performance improvement can be achieved 
by adapting them with respect to the phase (exploration or exploitation) of the 
learning process.

\section{Low-dimensional non-linear dynamical  systems}\label{sec:proof}
To demonstrate the usefulness of GPC for solving non-linear problems, GPC is 
first used on low-dimensional state-space systems exhibiting key non-linear 
features. The first one is an oscillator model displaying frequency 
cross-talk (section~\ref{ToC:desGMM}), the second one is a Lorenz system 
characterized by chaotic trajectories (section~\ref{ToC:desLor}). GPC is used to 
find control laws for both systems, either to achieve stabilization 
or to  optimize the Lyapunov exponents on the Lorenz 
system. These numerical systems have also served as playground for testing the 
influence of the different GPC parameters on convergence 
(section~\ref{sec:convergence}) and infer usable parameters for the experiments 
(section~\ref{sec:exp_parameters}). 

\subsection{Oscillator model}\label{ToC:desGMM}
In order to demonstrate nontrivial results in our control, we consider a simple system with 
frequency cross-talk
between a natural unstable oscillator and an actuated stable oscillator
via the base flow.
This model can be derived from the Navier-Stokes equation 
\cite{Luchtenburg2009jfm}
and has successfully described the actuation effect of
high and low-frequency forcing of flows around airfoils, bluff-bodies and of 
swirling jets.
It is arguably the most simple dynamical system prototype exhibiting frequency 
cross-talk. This model can be viewed as a generalization of the Landau model for 
the bifurcation from equilibrium to a periodic oscillation. A similar model can 
be built as a reduced-order model for a cylinder wake flow. As GPC is intended 
to be used on strongly non-linear systems where frequency cross-talk severely 
limits the use of a linear framework, we want to demonstrate the technique on the 
following model:
\begin{eqnarray}
\frac{\mathrm{d}}{\mathrm{d}t}\left[
\begin{array}{c}
a_1\\
a_2\\
a_3\\
a_4
\end{array}\right]
= \left[\begin{array}{cccc}
\sigma_1  & \omega_1 & 0 & 0 \\
-\omega_1 & \sigma_1 & 0 & 0 \\
0 & 0 & \sigma_2  & \omega_2 \\
0 & 0 & -\omega_2 & \sigma_2
\end{array}\right]
\left[
\begin{array}{c}
{a_1}\\
{a_2}\\
{a_3}\\
{a_4}
\end{array}\right]
+
\left[
\begin{array}{c}
0\\
0\\
0\\
b
\end{array}\right]\label{eq:GMM1}\\
\mbox{with }\sigma_1 = \sigma_{10} - (a_1^2 +a_2^2 + a_3^2 +a_4^2).\nonumber
\label{eq:osc_model}
\end{eqnarray}
Hereafter, we denote the sum of squared amplitudes as energy to avoid linguistic 
sophistication. We set $\omega_1=\omega_2/10=1$ and $\sigma_{10}=-\sigma_2=0.1$ 
so that the first oscillator $(a_1,a_2)$ is unstable at the origin while the 
other one $(a_3,a_4)$ is stable. When uncontrolled ($b\equiv 0$), the 
nonlinearity drives the first oscillator to nonlinear saturation by the change 
of total energy. The actuation  directly effects only the stable oscillator. We 
choose to stabilize the first oscillator around its fixed point $(0,0)$ and thus 
a cost function which measures the fluctuation energy of that unstable 
oscillator. For any useful application, the energy used for control is required 
to be small, hence, we penalize the actuation energy. The cost function is then 
defined as:
\begin{equation}
J=\left< a_1^2(t) + a_2^2(t) + \gamma b^2(t) \right>_T,
\label{eq:fitness}
\end{equation}
with $\gamma = 0.01$ as penalization coefficient and $\left<\cdot\right>_T$ 
denoting the time-average over the interval $[0, T]$. Here, 
$T=100\times{2\pi}/{\omega_1}$ is chosen to allow meaningful statistics. The 
quadratic form of the state and the actuation in the cost function is a standard 
choice in control theory. We apply GPC with full-state observation 
($\mathbf{s}\equiv\mathbf{a}$) to exploit all potential nonlinear actuation 
mechanisms stabilizing the first oscillator.

Knowing the nonlinearity at stake, an open-loop strategy can be designed: 
exciting the stable oscillator at frequency $\omega_2$ provokes an energy 
growth which stabilizes the first oscillator as soon as $a_3^2 +a_4^2 > 
\sigma_{10}$. Note that the linearization of~\eqref{eq:GMM1} yields two 
uncoupled oscillators. Thus, the first oscillator is uncontrollable in a linear 
framework. 
\subsubsection{Results}

The function space is explored by using a set of elementary ($+,-,\times,/$) and 
transcendental ($\exp,\,\sin,\,\ln$ and $\tanh$) functions. The functions are 
'protected'  to allow them to take arbitrary arguments in $\mathbb{R}$ (e.g. a 
thresholding is achieved on denominators in divisions to avoid division by 
zero). Additionally, the actuation command is limited to the range $[-1\,,\, 1]$ 
to emulate an experimental actuator. Up to $50$ generations comprising $1000$ 
individuals  are processed. The tournament size is $n_p=7$, elitism is set to 
$n_e=1$, the probabilities of replication, crossover and mutation are  
$P_r=0.1$, $P_c=0.6$ and $P_m=0.3$ respectively (see Tab.~\ref{tab:GMMparam}). 
\begin{table}
{
\caption{GPC parameters used for the control of model~\eqref{eq:osc_model}.}
\label{tab:GMMparam}
\begin{tabular}{cc}
\hline\hline
Parameter & Value\\\hline
n & 1000\\
$P_r$ & 0.1\\
$P_m$ & 0.3\\
$P_c$ & 0.6\\
$n_p$ & 7\\
$n_e$ & 1\\
node functions & $+,-,\times,/,\sin,\exp,\log,\tanh$\\
\hline\hline
\end{tabular}
}
\end{table}

\begin{figure}
\centerline{\includegraphics[width=0.4\textwidth,natwidth=610,natheight=692]
{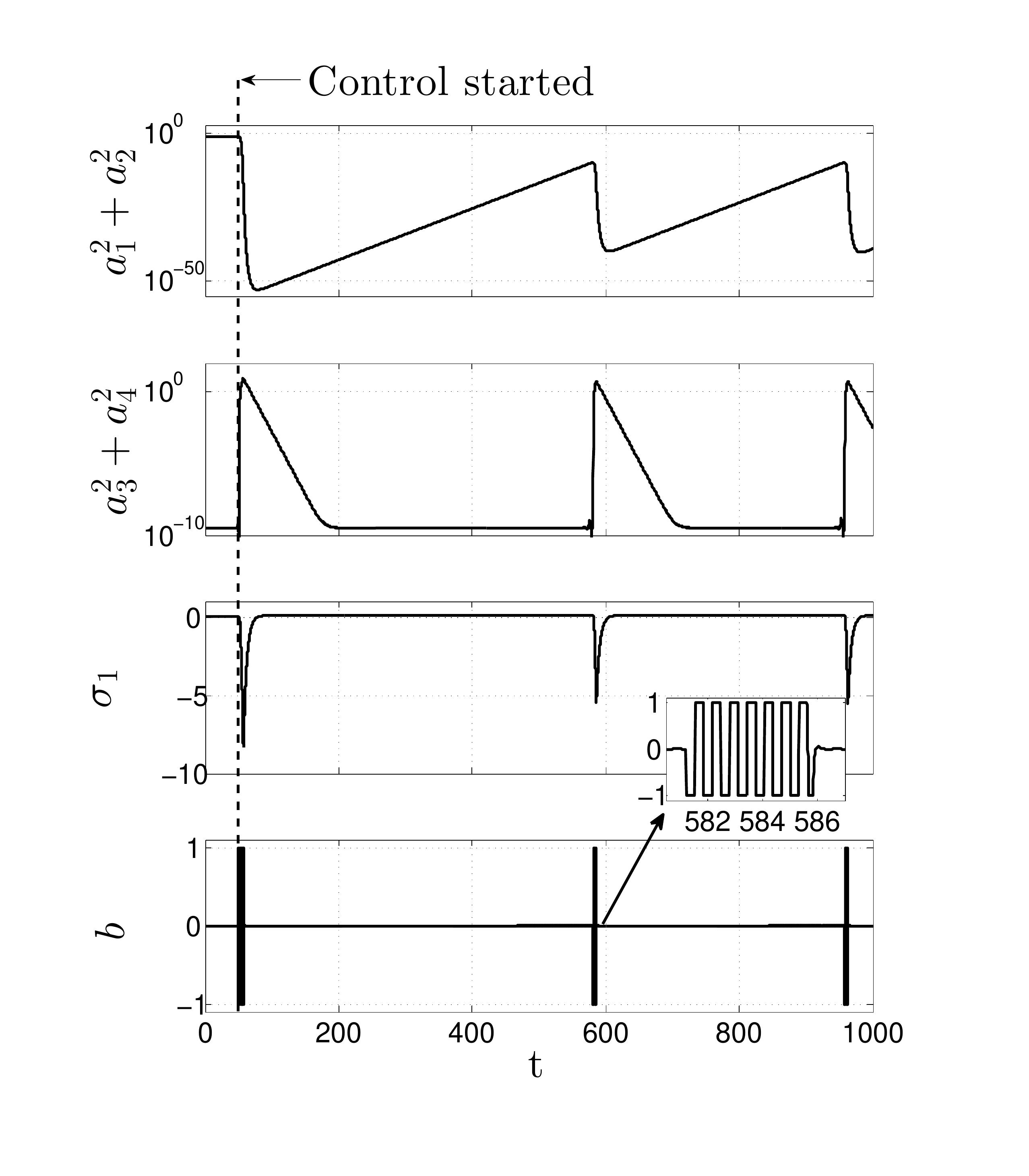}}
\caption{Genetic programming control of the model. When the energy 
contained in the first oscillator (top) is larger than $10^{-10}$ the control 
(bottom) is exciting the second oscillator at frequency $\omega_2$, its energy 
grows so that $\sigma_1$ reaches approximately $-5$. This results in a fast 
decay of the energy in the first oscillator after which the control goes in 
``standby mode``. An animation of the controlled system can be found in 
supplemental video S1.}
\label{fig:energies}
\end{figure}

\begin{figure}
\includegraphics[width=0.4\textwidth,natwidth=610,natheight=692]{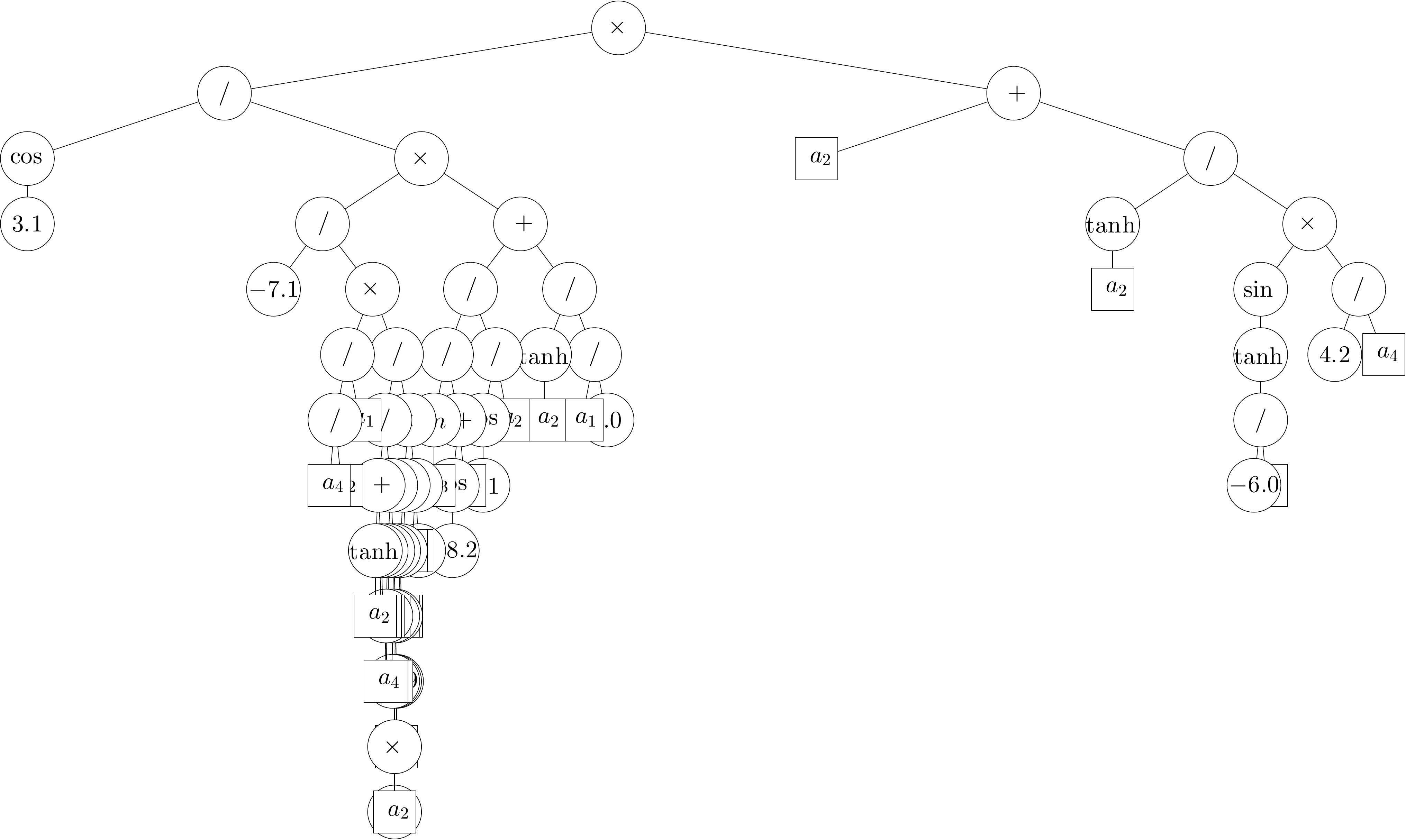}
\caption{Tree representation of the best individual obtained by GPC for the 
control of model~\eqref{eq:osc_model}.\label{GMMtree}}
\end{figure}

The control law ultimately returned by the GPC process corresponds to the best 
individual of the last generation. The performance and the behavior of the 
control law are displayed in Fig.~\ref{fig:energies}. This control law is 
energizing the second oscillator up to $10^0 \gg \sigma_{10}$ as soon as the 
first oscillator has an energy which is larger than $10^{-10}$. This is 
stabilizing the unstable oscillator very quickly with an amplitude decay scaling 
as $10^{-0.1t}$. After stabilization, the control amplitude stays at very low 
values. That keeps the stable oscillator at a correspondingly low energy 
($\approx 10^{-10}$), 
while the amplitude of the unstable oscillator is exponentially increasing with 
its initial growth rate $\sigma_{10}$. This control law exploits the frequency 
cross-talk and vanishes when not needed, i.e. $a_1\approx a_2\approx 0$. The 
expression tree of the best solution found by GPC can be viewed in 
Fig.~\ref{GMMtree}. It can be summarized as follows:
\begin{multline}
b=K_1(a_4)\times K_2(a_1,a_2,a_3,a_4)\quad\text{with}\\
\shoveleft{K_1(a_4)=5.475 \times a_4}\quad \text{and}\\
\shoveleft{K_2(a_1,a_2,a_3,a_4)= \frac{(\frac{(\frac{a_4}{4.245}) \times 
(\sin(\tanh(\frac{a_4}{-5.987})))}{\tanh(a_2)} + a_2)}{\cos(3.053)} \times}\\
\shoveright{(\frac{\frac{8.965}{a_1}}{\tanh(a_2)} + 
\frac{\frac{a_2}{\cos(3.053)}}{\frac{a_4 + \cos(-8.208)}{\log(a_3)}}) 
\times\\(\frac{(\frac{(\frac{a_1}{\frac{a_1}{a_4}})}{\frac{(\frac{(\sin(a_4)) 
\times (\tanh((4.640) \times (a_2)))}{-6.912 - (a_4)}) \times (\frac{a_1}{(a_2) 
\times (a_4)})}{\frac{a_2}{a_1} + \tanh(a_2)}}) \times 
(\frac{a_1}{\frac{a_2}{a_4}})}{-7.092}).}
\end{multline}
The function $K_1(a_4)$ describes a phasor control that destabilizes the stable 
oscillator. The function $K_2(a_1,a_2,a_3,a_4)$ acts as a gain dominated by the 
energy of the unstable oscillator. That control could not be derived from a 
linearized model of the system. Moreover, less energy is used as compared to the 
best periodic excitation.

\subsection{Lorenz system}\label{ToC:desLor}
One well-known feature of non-linear systems with dimension larger than three 
is their ability to display 
chaotic behavior. The control of such systems is a field in its own. Chaos being 
certainly unfavorable to many applications, many studies have investigated the 
possibility to stabilize chaotic trajectories in periodic 
orbits~\citep{ott1990controlling,Pyragas1992,schusterhandbook}. On the other 
hand, chaos can be sought in order to increase mixing~\cite{stroock2002chaotic}. 
We use GPC on a Lorenz system controlled in the third component in order to 
optimize the maximal Lyapunov exponent while keeping the system in a finite 
state space.

\subsubsection{Problem formulation}
The system is formulated as follows:
\begin{equation}
\frac{\mathrm{d}a_1}{\mathrm{d}t} =  \sigma\left(a_2 - a_1\right),\quad
\frac{\mathrm{d}a_2}{\mathrm{d}t} = a_1\left(\rho-a_3\right)-a_2,\quad
\frac{\mathrm{d}a_3}{\mathrm{d}t} =  a_1a_2 - \beta a_3 +b
\label{eq:Lorenz} 
\end{equation}
with full-state feedback $b=K(a_1, a_2, a_3)$, i.e.\ 
$\mathbf{s}\equiv\mathbf{a}$. With $\sigma=10$ and $\beta=8/3$, the Lorenz 
system can be stable, periodic or chaotic depending on the value of $\rho$. We 
employ $\rho=20$, such that the uncontrolled system ($b\equiv0$) is periodic. 
Like \cite{de2014optimizing},  we aim at maximizing the largest Lyapunov 
exponent $\lambda_{1}$ while penalizing the actuation power with a factor 
$\gamma$. If $\lambda_{1}$ is positive, the system is chaotic and well-mixing. 
We define the cost function, which should be minimized, as: 
\begin{equation}
\begin{array}{l l l l}
J   & =           & \exp(-\lambda_{1}) +\gamma\left< b^2(t)\right>_T  
&\quad\mbox{if}\,\, \sum_{i=1}^3 \lambda_i < 0,\\
J  &\rightarrow & \infty&\quad\mbox{if}\,\, \sum_{i=1}^3 \lambda_i \geq 0,
\end{array}
\end{equation}
where $T=100$ is the integration time and $\lambda_1\ge\lambda_2\ge\lambda_3$ 
are the {finite-time} Lyapunov exponents (FTLE). 
These exponents are obtained {by calculating the local Jacobian for each time 
step and integrating its eigenvalues along the trajectory}. 
$J$ is assigned the largest computable real number on the computer 
if the sum of the Lyapunov exponents is positive or the states exceed the bounds 
we specify. 
\subsubsection{Results}
\begin{figure}
\begin{center}
\begin{tabular}{c c}
\includegraphics[width=0.35\textheight,natwidth=610,natheight=692]
{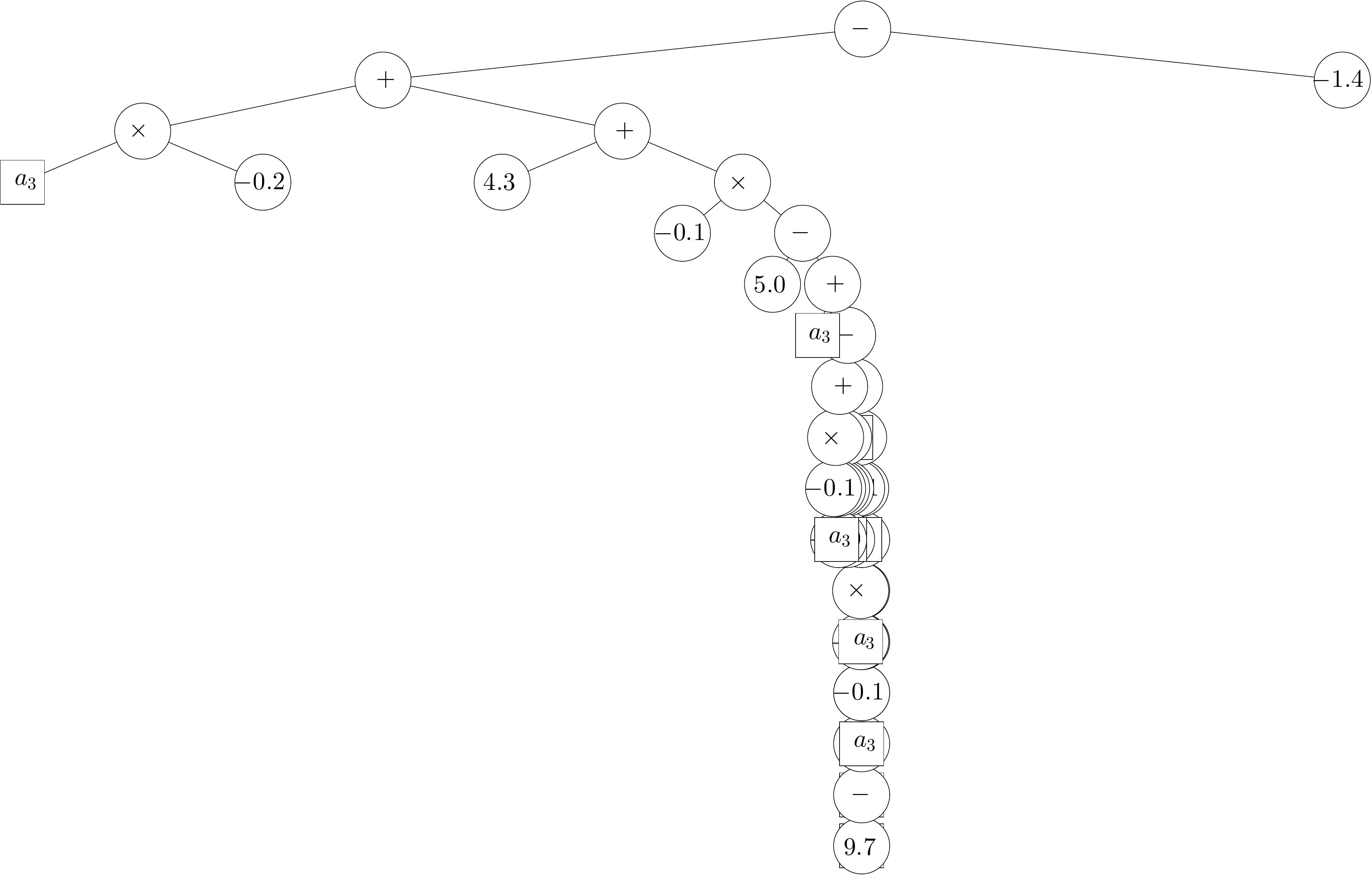}&a)\\
\includegraphics[width=0.35\textheight,natwidth=610,natheight=692]
{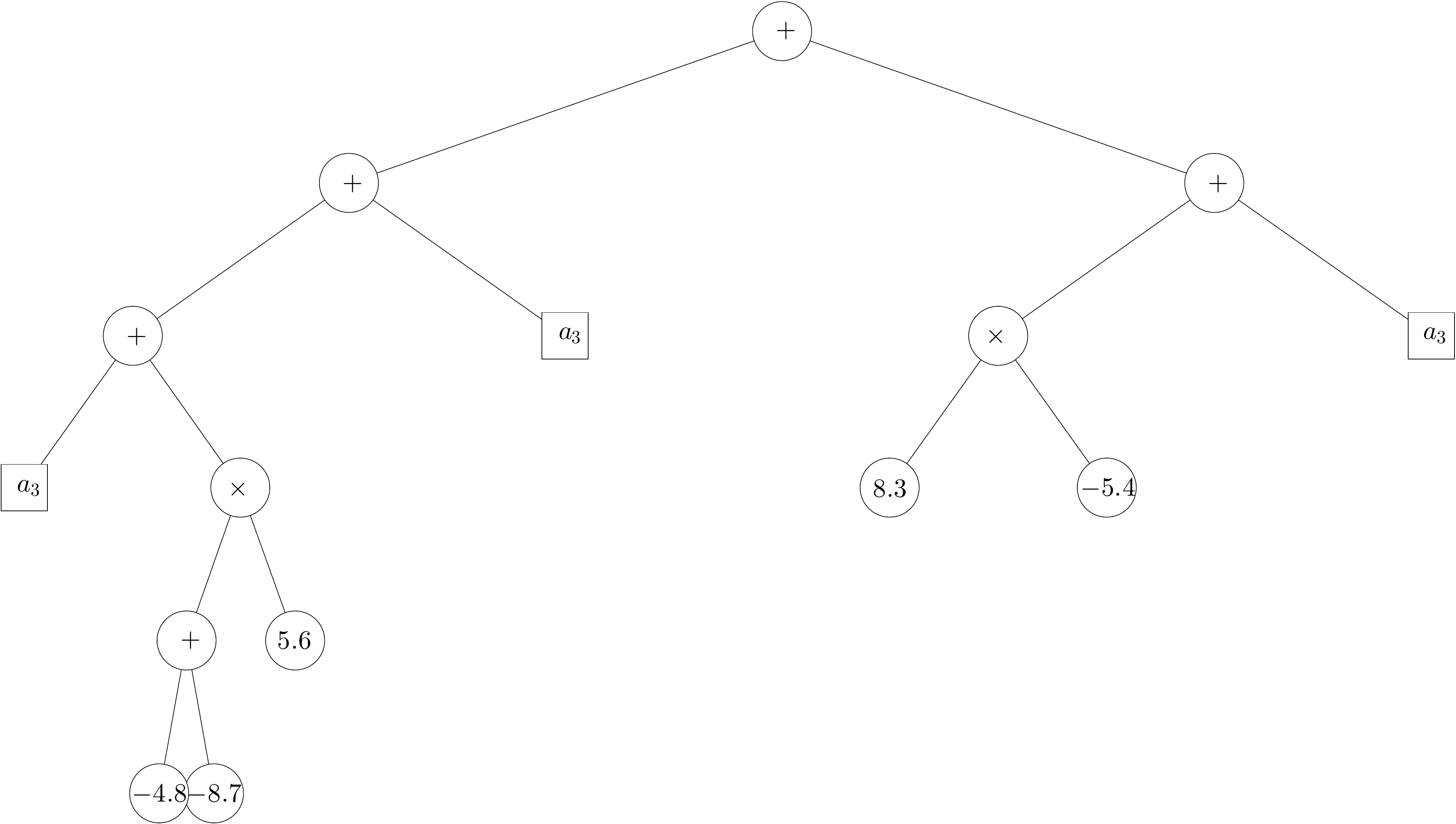}&b)\\
\includegraphics[width=0.35\textheight,natwidth=610,natheight=692]
{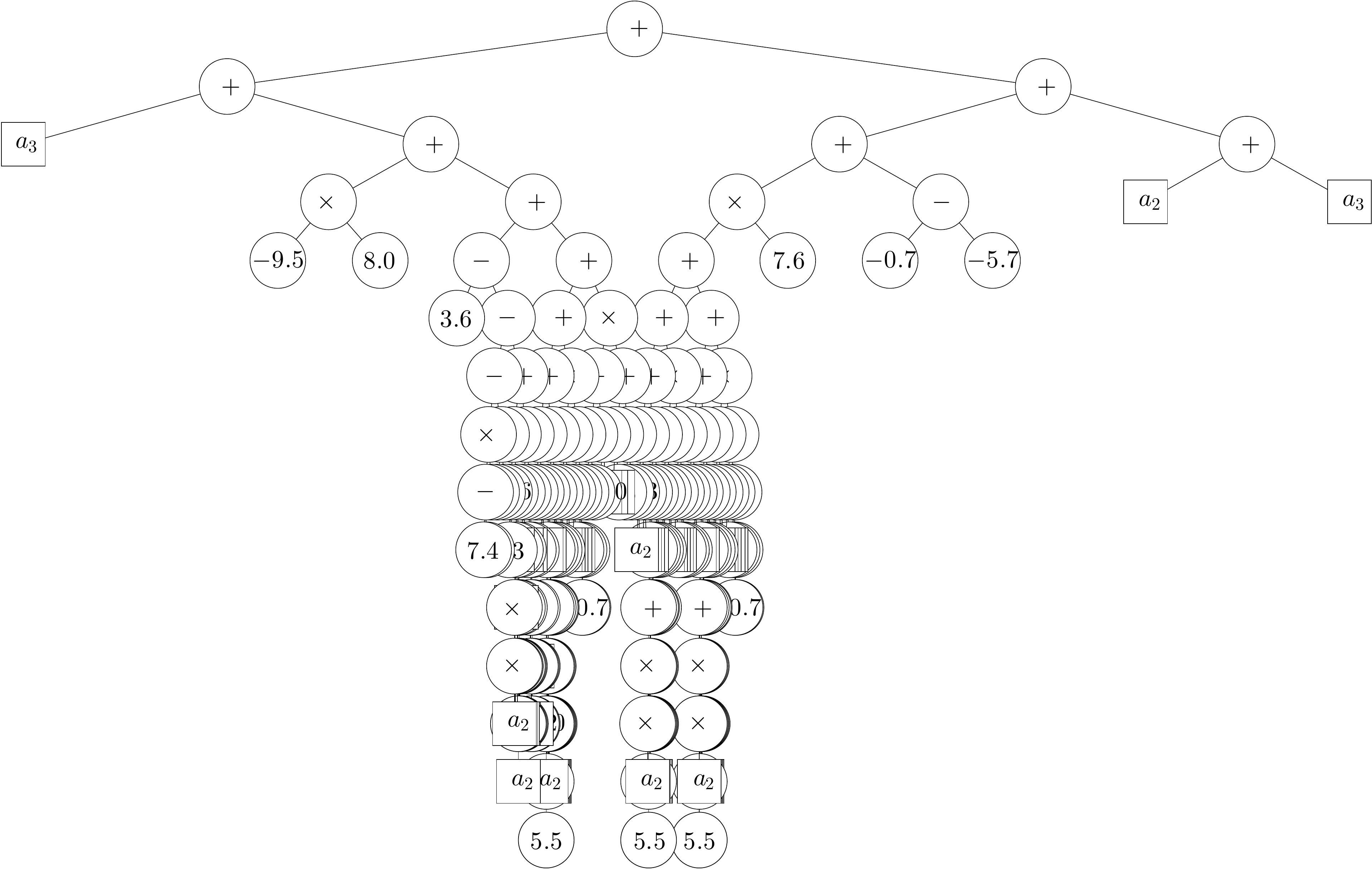}&c)
\end{tabular}
\end{center}
\caption{Tree representation of the best individuals obtained by GPC for the 
control of the Lorenz system with $\gamma=1$ (a), $\gamma=0.01$ (b) and 
$\gamma=0$ (c).\label{gamma001tree}}
\end{figure}
\begin{figure}
\begin{center}
\includegraphics[width=0.4\textwidth,natwidth=610,natheight=692]{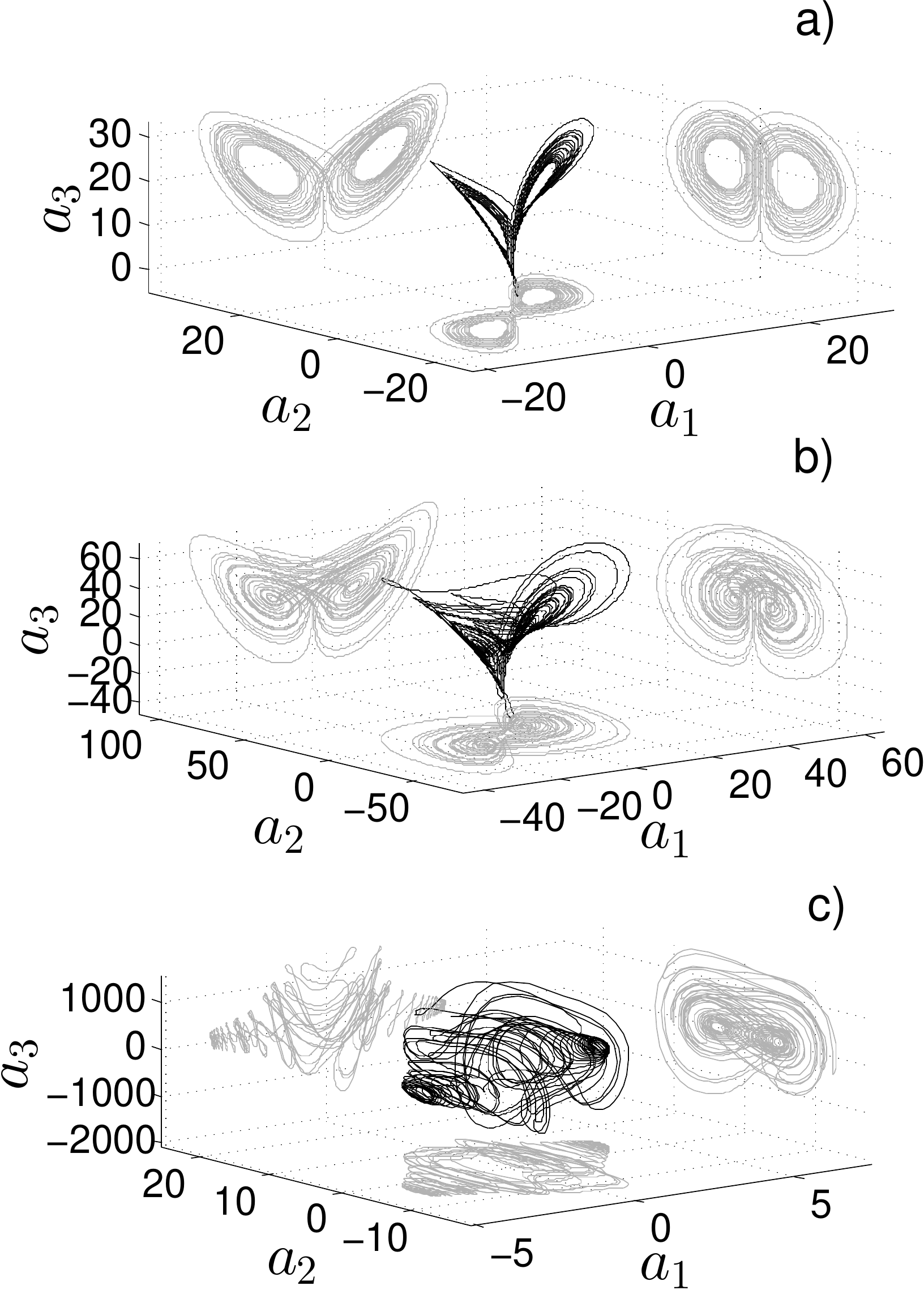}
\end{center}
\caption{Controlled Lorenz systems with $\sigma=10$, $\beta=8/3$ and $\rho=20$. 
For $\gamma=1$ (a), the system exhibits chaotic behavior ($\lambda_{1}=0.715$) 
close to the canonical chaotic Lorenz attractor with $\rho=28$ 
($\lambda_{1}=0.905$). For $\gamma=0.01$ (b), the system exhibits more complex 
trajectories, the nature of the central fixed point has changed and 
$\lambda_{1}=2.072$. For $\gamma=0$ (c), the nature of all fixed points has 
changed. The non-penalization of the actuation leads to a change in the scales 
($\lambda_{1}=17.613$). An animation of the controlled system can be found in 
supplemental video S2.}
\label{fig:3DLorenz}
\end{figure}

GPC is applied to the periodic Lorenz system to maximize the largest 
{finite-time} Lyapunov exponent while keeping the solution bounded. The basic 
operations that compose the control law are ($+$, $-$, $\times$, $/$) as well as 
randomly generated constants. The maximum number of generations is chosen as 50 
with 1000 individuals each. {The tournament size is $n_p=7$, elitism is set to 
$n_e=1$, the probabilities of replication, crossover and mutation are 
respectively $P_r=0.1$, $P_c=0.6$ and $P_m=0.3$ (see 
Tab.~\ref{tab:lorenzparam}).}
\begin{table}
{
\caption{GPC parameters used for the control of the Lorenz 
system}\label{tab:lorenzparam}
\begin{tabular}{cc}
\hline\hline
Parameter & Value\\\hline
n & 1000\\
$P_r$ & 0.1\\
$P_m$ & 0.3\\
$P_c$ & 0.6\\
$n_p$ & 7\\
$n_e$ & 1\\
node functions & $+,-,\times,/,\sin,\exp,\log,\tanh$\\
\hline\hline
\end{tabular}
}
\end{table}
We consider for $\gamma$ the values of $\gamma_S=1$, $\gamma_W=0.01$ and 
$\gamma_N=0$, representing strong, weak and no penalization of the actuation.
This illustrates how the cost function definition influences the problem to be 
solved. 
After 50 generations, the best individuals (see Fig.~\ref{gamma001tree}) 
associated with strong, weak and no penalization have maximum FTLE of 
$\lambda_{1}=0.715$, $2.072$ and $17.613$, respectively. The changes in the 
system and the control function are displayed in Fig.~\ref{fig:3DLorenz}. The 
control laws associated with $\gamma_S$ and $\gamma_W$ are affine expressions of 
$a_3$ and the reduction of the actuation cost leads to a larger amplitude of the 
feedback. In those cases, the most efficient control leads the system into 
behaviors close to the canonical Lorenz system ($\rho=28$, $\lambda_{1}=0.905$). 
For $\gamma_W$ the nature (from saddle point to spiral saddle point) and the 
position of the central fixed point from the actuated system are changed. If the 
actuation is not penalized ($\gamma=0$) the feedback law is a complex nonlinear 
function of all states. The nature and position of all fixed points are changed 
as $\lambda_{1}$ reaches higher values. 
or model-free approach has been proposed in the literature to optimize the 
largest Lyapunov exponent.

\subsection{Convergence}\label{sec:convergence}
\begin{figure*}
\includegraphics[width=0.9\textwidth]{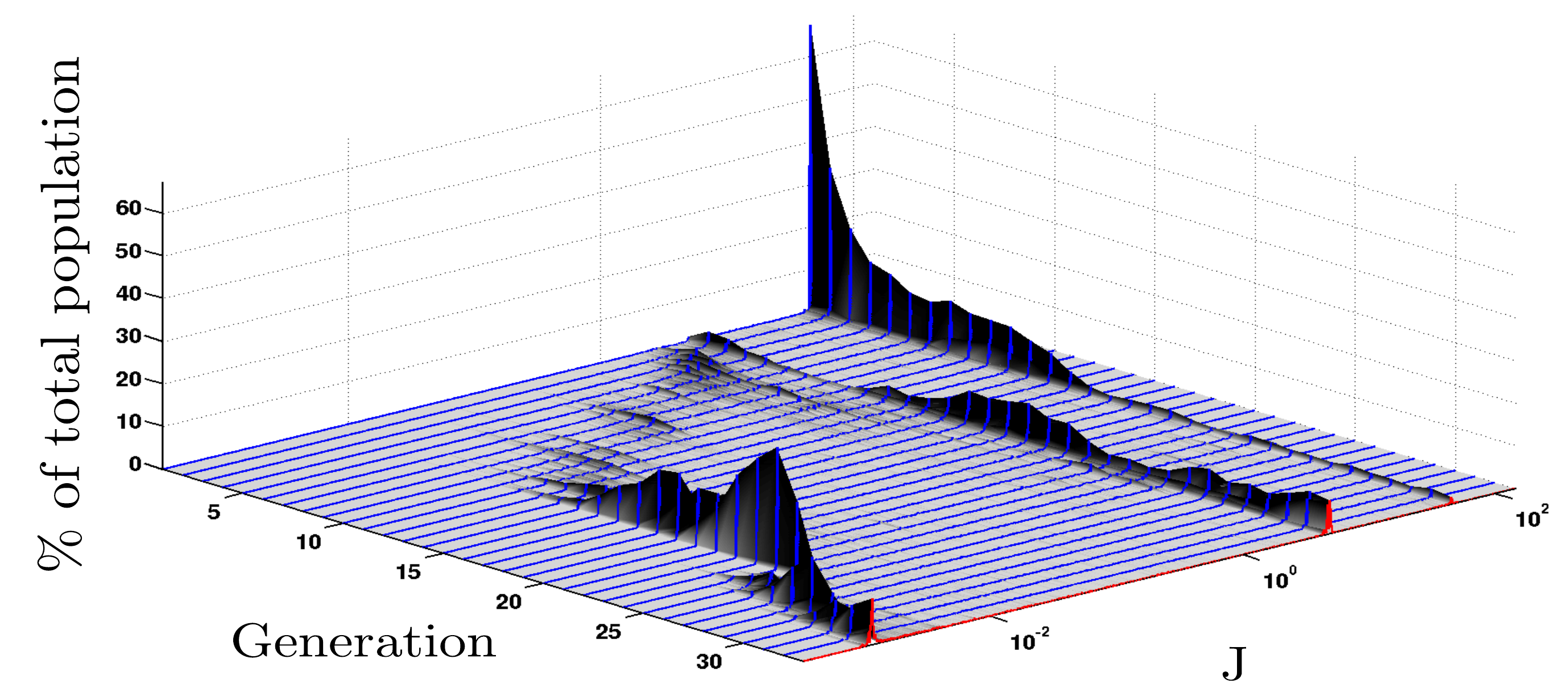}
\caption{Successive histograms of population repartition with respect to the 
cost function values for the stabilization of the oscillator model. 
High peaks indicate the high density of individuals around the cost function 
value. The histograms use $1000$ bins which are logarithmically distributed 
between the first non-nul value and the maximum value of the cost function 
evaluations.}\label{fig:convergence}
\end{figure*}
Both examples illustrate how GPC is progressing towards the minimum of the cost 
function. The statistical process that selects the individual for breeding 
allows individuals which are not optimal to be selected. This keeps diversity in 
the population and ensures that the GPC process is not confined in a local 
minimum. In figure~\ref{fig:convergence} successive histograms of the 
repartition of cost function values in the population are plotted for the 
model~\eqref{eq:osc_model}. The data comes from section~\ref{ToC:desGMM} but 
the aspect of the graph is typical of any evolutionary algorithm search. The 
population is stirred towards successive local minima. Around these minima an 
increasing number of individuals is exploring the search-space leading in a 
local repartition of cost function values due to the effect of crossovers. 
Mutations are mostly responsible of the jumps between local minima. Once a 
better local minimum is found, a large part of the population is shifted in that 
direction while other local minima are iteratively depleted. While a large part 
of the population lives around the minima, smaller ripples indicate the still 
existing diversity in the population. The two large \emph{families} of 
individuals existing throughout the whole process account for obvious behaviors: 
individuals with null output (generated by a multiplication by $0$ obtained by 
the difference of two identical subtrees e.g.) or with constant and saturated 
output. For the stabilization of our oscillator model, the GPC stopped after 35 
generations, both oscillator energy and control energy vanished below  numerical 
accuracy of the integration scheme (not represented in 
Fig.~\ref{fig:convergence} due to the logarithmic scale). 

\subsection{Conclusions on usability of GPC and best practices for 
            experiments}\label{sec:exp_parameters}
The stabilization of the simple oscillator model has shown that GPC is able 
to exploit strongly non-linear mechanisms to solve a control problem. The 
solution found is explicitly exciting one oscillator at his eigenfrequency to 
stabilize another oscillator at another frequency. This could not be achieved in 
a linear framework without the knowledge of the existing non-linearity. The 
chaotization of the Lorenz system highlights the fact that GPC can minimize a 
complex cost functional associated with a non-linear system. These GPC 
realizations on low-order systems allow to guide the parameters choices before 
applying GPC to experiments. The two main differences between these numerical 
test-cases and the experiments that will be presented in the next section are 
(i) the fact that the experiments are not parallelized and (ii) the fact that 
the cost function values are obtained with a measurement uncertainty. The total 
time of experiment being limited by the operation cost and facilities 
availability, limiting the number of individuals by generation is necessary. As 
for the second point, the uncertainty can be reduced (but not nullified) by 
increasing the evaluation time, which is not desired as each additional second 
of evaluation can easily translate in additional hours of total experimental 
time. This is partially solved by the re-evaluation process introduced in 
section~\ref{sec:GP:Evaluation}. Nonetheless, an intrinsic noise on the experiment 
can never be avoided. This means that the intrinsic uncertainty level on the 
evaluation of the cost function is the utmost limit that can be reached while 
looking for minima of said cost function. This devalues the higher need of 
crossovers defended by GP experts, while a higher rate of mutation can alleviate 
the need of exploration and diversity induced by a reduced population. Hence, 
the parameters that will be used in experiments will be shifted towards high 
mutation rate and reduced populations.

\begin{figure*}[ht]
\begin{center}
\includegraphics[width=0.70\textwidth]{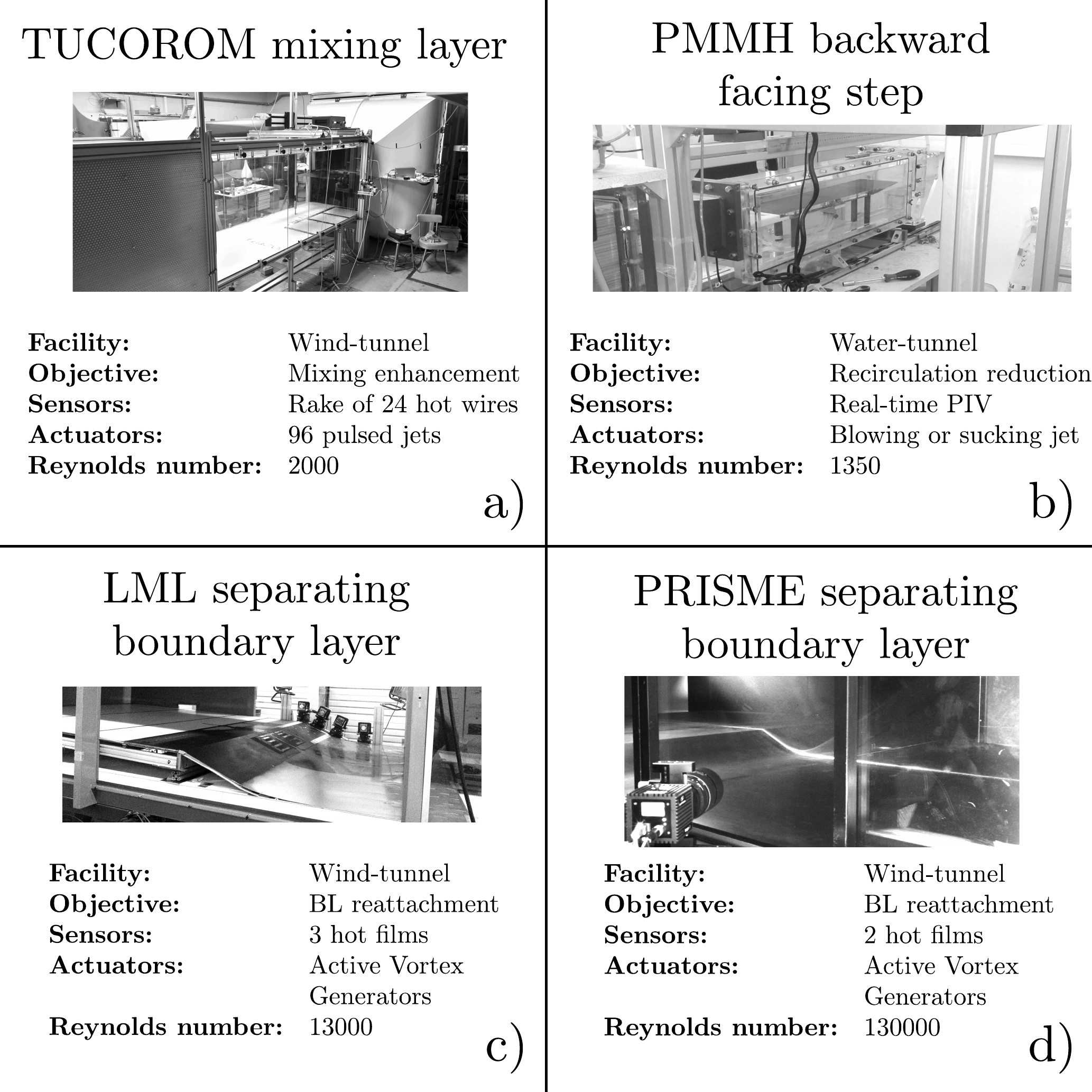}
\end{center}
\caption{Picture and global flow control parameters of the four experiments. a: TUCOROM mixing layer. b: PMMH backward facing step. c: LML separating boundary layer. d: PRISME separating boundary layer. }\label{fig:exp_id}
\end{figure*}
\section{Experimental demonstrators}\label{sec:exp}
GPC has been applied to four experimental demonstrators: in the TUCOROM mixing 
layer (section~\ref{sec:TUC}), in the PMMH backward facing step flow 
(section~\ref{sec:PMMH}), in the LML separated boundary layer 
(section~\ref{sec:LML}) and in the PRISME separated boundary layer 
(section~\ref{sec:PRISME}). These four experiments are presented in 
Fig.~\ref{fig:exp_id}.

\subsection{Control of a turbulent mixing layer}\label{sec:TUC}
The first experimental implementation of GPC has been achieved in the TUCOROM 
mixing layer demonstrator (Fig.~\ref{fig:exp_id}a). The goal of the experiment 
was to enhance the mixing properties of the mixing layer.

\subsubsection{Experimental setup}
\begin{figure}
\begin{center}
\includegraphics[width=0.4\textwidth]{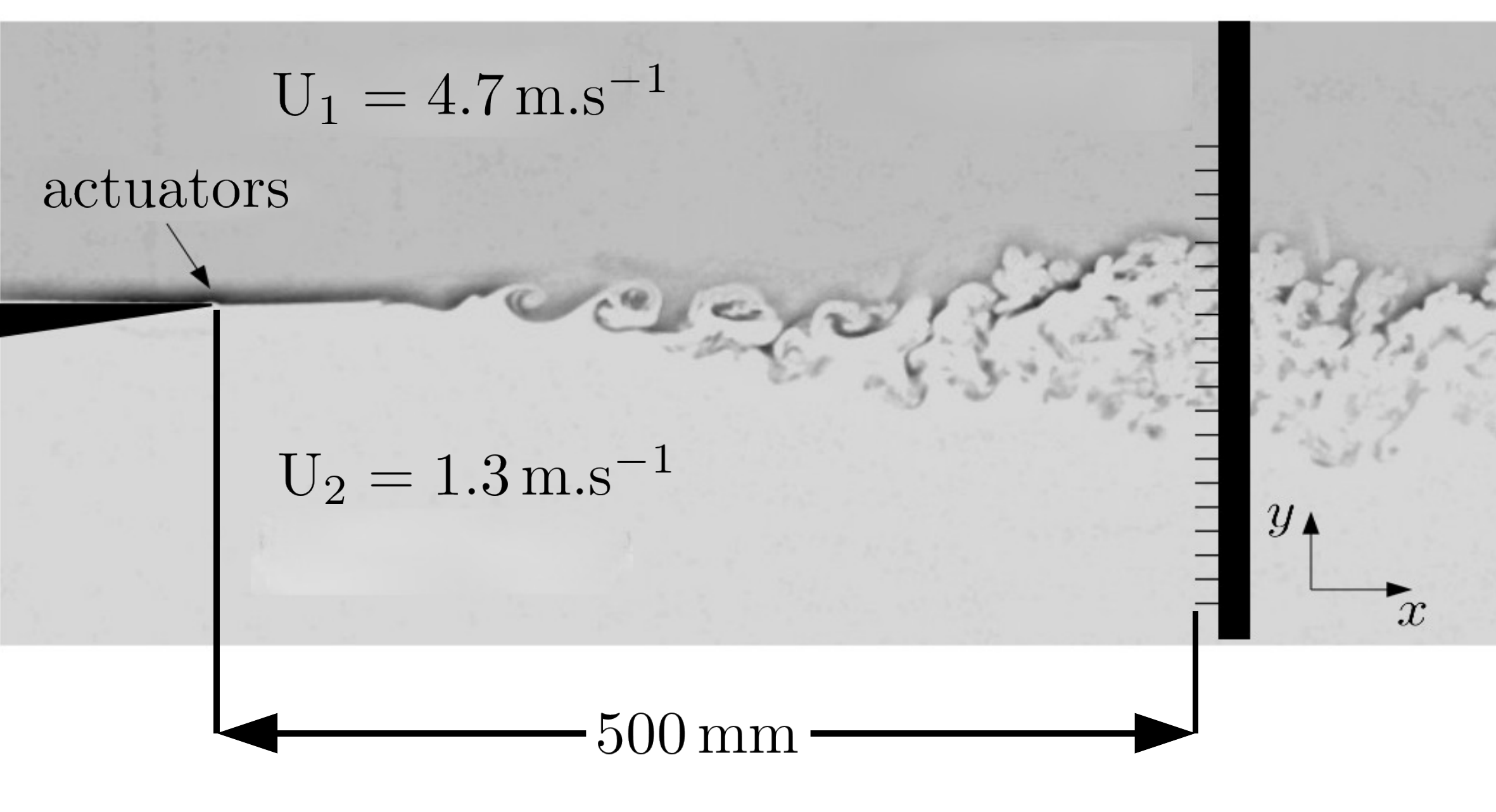}
\end{center}
\caption{Experimental configuration of the TUCOROM mixing layer.
The hot-wire rake is placed $500\,mm$ downstream of the separating plate
to capture the structures in the shear layer.
The vertical spacing of the hot-wire probe is $\Delta{y}=8\,mm$.\label{fig:expsetup}}
\end{figure}
The TUCOROM  mixing layer demonstrator~\cite{Parezanovic2015ftac} consists in a 
dual turbine wind tunnel which allows to set two different velocities on each 
part of a splitter plate with a velocity ratio $U_1/U_2=3.6$ and Reynolds number 
based on the initial mixing layer thickness between $500$ (laminar) and $2000$ 
(turbulent). The test-section is $3\,\text{m}$ long with a subsection of 
$1\times 1\,\text{m}^2$. Inside the splitter plate 96 micro-jets allow to blow 
in the streamwise direction. A rake of 24 hot wires, simultaneously acquired at 
$20\, \text{kHz}$, is placed at $500$ mm downstream of the splitter plate. The 
sensors used for the genetic programming are the velocity fluctuations for 9 
chosen, equi-distributed, sensors across the shear layer velocity gradient, 
while all 24 sensors are used for the evaluation of a given control law. GPC is 
applied on the configuration described in Fig.~\ref{fig:expsetup}b with the 
following cost function $J$:
\begin{equation}
J=\frac{1}{W},\,\, \mbox{with}\,\, W=\frac{\left<\left[\sum_{i=1}^{24}{s'}_i^2(t)\right]\right>_T}{\mbox{max}_{i \in [1,24]}(\left<{s'}_i^2\right>_T)},\label{eq:Jexp}
\end{equation}
where $\left<\cdot\right>_T$ is the average over the evaluation time 
$T=10\,\text{s}$, $s_i$ represents the hot wire signal from hot wire $i$, and 
$s_i'$ its fluctuation calculated over $T/10$. $W$ can be interpreted as the 
width of the fluctuation profile at the considered position. GPC is applied with 
$\left(+, -, \times, /, \sin, \cos,\exp, \tanh, \log\right)$ as node functions 
for the expression-trees. The parameters used are summarized in 
Tab.~\ref{tab:chexparam}.

\begin{table}
{
\caption{GPC parameters used for the control of the TUCOROM mixing layer }\label{tab:chexparam}
\begin{tabular}{cc}
\hline\hline
Parameter & Value\\\hline
\multirow{2}{*}{n} & 1000 (first generation)\\
 & 100 (other generations)\\
$P_r$ & 0.1\\
$P_m$ & 0.25\\
$P_c$ & 0.65\\
$n_p$ & 7\\
$n_e$ & 1\\
Node functions & $+,-,\times,/,\sin,\cos,\exp,\log,\tanh$\\
\hline\hline
\end{tabular}
}
\end{table}

\subsubsection{Results}
\begin{figure}
\begin{center}
\includegraphics[width=0.4\textwidth]{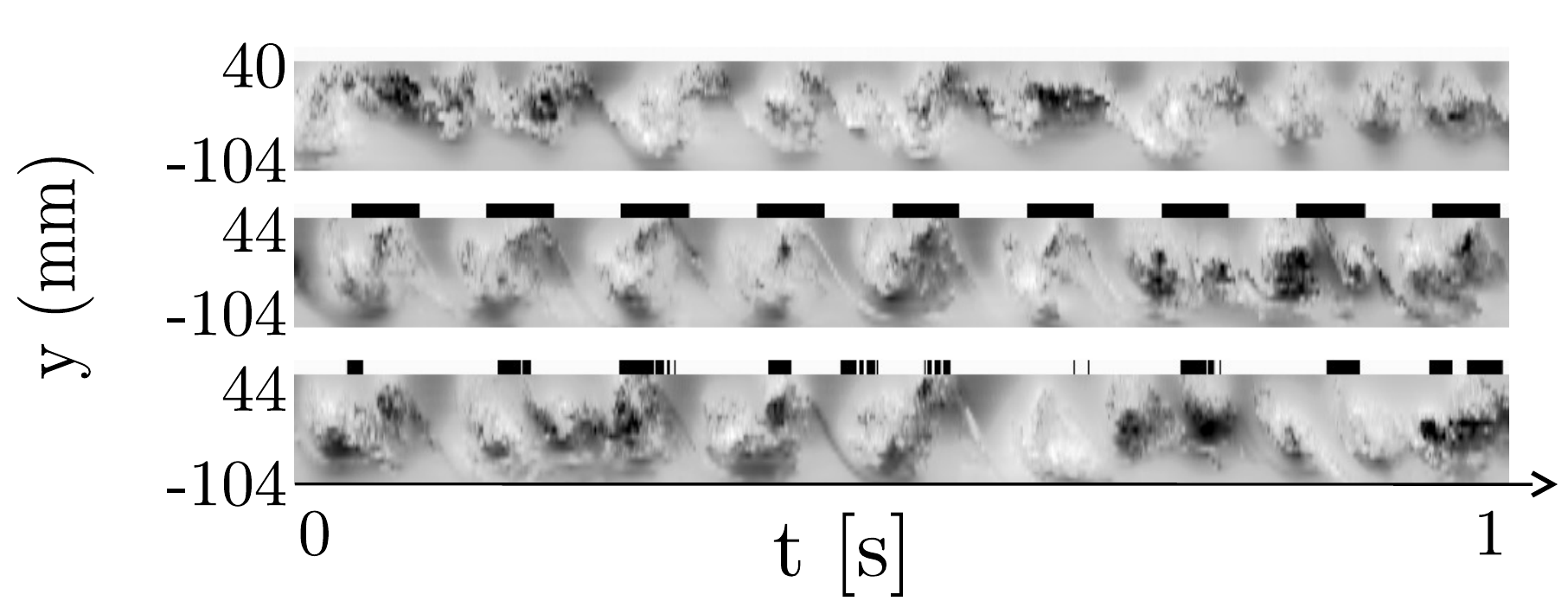}
\end{center}
\caption{Pseudo-visualizations of the TUCOROM experimental mixing layer demonstrator~\cite{Parezanovic2015ftac}
for three cases: natural baseline (top, width $W=100$ \%),
the best open-loop benchmark (middle, width $W=154$ \%),
and GPC closed-loop control (bottom, width $W=170$ \%).
The velocity fluctuations recorded  by 24 hot-wires probes (see Fig.~\protect\ref{fig:expsetup})
are shown as contour-plot over the time $t$ (abscissa)
and the sensor position $y$ (ordinate).
The black stripes above the controlled cases indicate
when the actuator is active (taking into account the convective time).
The average actuation frequency achieved by the GPC control is comparable to the open-loop benchmark.}
\label{fig:chex}
\end{figure}

Many parameters for both the experiment and GPC have been 
tested~\cite{parezanovic2015pof}. We display the results for the configuration 
of Fig.~\ref{fig:expsetup} at the Reynolds number of $500$ with the cost 
functional~\eqref{eq:Jexp}. Figure~\ref{fig:chex} shows the pseudo 
visualizations obtained for the natural flow, the best open-loop control (an 
harmonic forcing at frequency $f_a\approx 9\, \text{Hz}$ and duty cycle 50 \%) 
and the GPC obtained control. Table~\ref{tab:chex} displays the corresponding 
cost function values obtained for these three cases.

\begin{table}
{
\caption{Performance of uncontrolled, best open-loop controlled and GPC controlled systems. GPC manages to increase the width of the shear layer by $70\%$, outperforming the best open-loop control by $16\%$. Additionally, the actuation cost is $48\%$ of the best open-loop control.}\label{tab:chex}
\begin{tabular}{cccc}
\hline\hline
case & natural & best open-loop & GPC\\\hline
J & $1$ & $0.65$ & $0.59$\\
W & $100\%$ & $154\%$ & $170\%$\\
Actuation & \multirow{2}{*}{0} & \multirow{2}{*}{1} & \multirow{2}{*}{0.48}\\
Cost & & &\\
\hline\hline
\end{tabular}
}
\end{table}

GPC yields a 70 \% increase of the mixing layer width, which outperforms open-loop forcing by 16 \%. Furthermore the actuation cost is $48\%$ of what is used for the open-loop case (Tab.~\ref{tab:chex}). This experiment also demonstrates that control laws obtained at different operating conditions are still performing, though not optimally, while open-loop controls can only be applied to given operating conditions. This robustness is directly due to the native retro-action implemented in closed-loop control. Further improvement could be achieved by specifying the desired robustness inside the cost functional.

\subsection{Control of a separated flow using Real-Time PIV}\label{sec:PMMH}
GPC has been applied in the PMMH water tunnel (Fig.~\ref{fig:exp_id}b) on a separated flow over a backward facing step with the goal to reduce the recirculation zone.
\subsubsection{Experimental setup}
\begin{figure}
\centering
\includegraphics[width=0.4\textwidth]{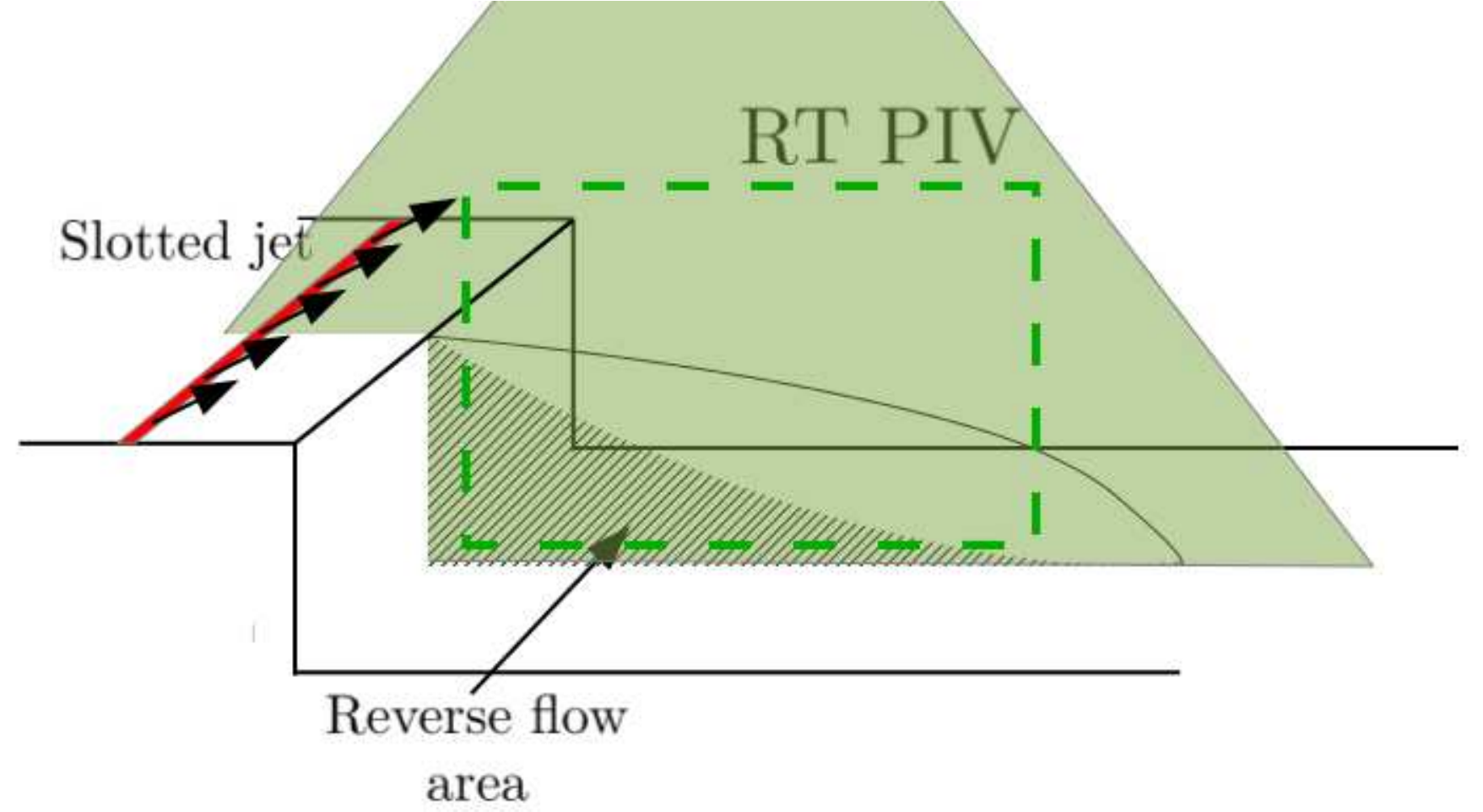}
\caption{Experimental configuration of the flow over a backward facing step in the PMMH water tunnel. A slotted jet is situated just upstream of the separation allowing to achieve blowing or suction in the boundary layer. Real-Time PIV is achieved using a laser sheet in the symmetry plane. This allows to detect the reversed flow region (striped area).}
\label{fig:PMMH_expsetup}
\end{figure}

The PMMH water tunnel is gravity driven, allowing to reach speeds up to $22\,\text{cm.s}^{-1}$. In the current experiment (Fig.~\ref{fig:PMMH_expsetup}b) a backward facing step of height $h=1.5\, \text{cm}$ is placed in the $L\times l\times H = 80\times15\times20\, \text{cm}^3$ test-section. The Reynolds number, based on the freestream velocity $U_{\infty} = 7.3\,\text{cm.s}^{-1}$,  is $Re_h={U_\infty h}/{\nu}=1350$. Actuation is achieved thanks to a slotted jet located at a distance $2h$ upstream the step edge and can perform either blowing or suction. The angle between the jet and the wall is $45$\textdegree\ in the direction of the flow. A real-time PIV system~\cite{Gautier2013OF,Gautier2013control} is used to compute flow fields at $42\,\text{Hz}$. The sensor signal used for GPC is the ratio between controlled and natural recirculation area of the flow obtained through instantaneous PIV fields, i.e.:
\begin{eqnarray}
s(t)&=&\frac{A(t)}{A_u},
\end{eqnarray}
where
\begin{eqnarray}
A(t)&=&\int H(-u(t))(x,y)\hspace{1mm}\mathrm{d}x\mathrm{d}y,\\
A_u&=& \left< A(t) \right>,\, \text{without actuation}.
\label{eq:A_r}
\end{eqnarray}
Here, $u$ is the streamwise velocity component, $H$ the Heaviside function and $\left<\cdot\right>_T$ a time averaged value of its argument. Note that $A_u$ is the time-averaged recirculation area for the uncontrolled flow.
The cost function is aimed at minimizing the back-flow region and then the recirculation area, while penalizing the actuation cost:
\begin{equation}
{\displaystyle J=\langle s \rangle_T + \gamma \langle |b| \rangle_T^2},
\label{eq:cost_function}
\end{equation}
with $b$ the control command (a signed value proportional to the flow rate 
through the jet) and $\gamma$ a penalization coefficient. Although the choice 
for $\gamma$ is arbitrary, the value represents a trade-off between the gain on 
the area reduction and the cost of actuation. Setting a low (respectively high) 
value of $\gamma$ means that the performance of the system is much more 
(respectively less) important than the cost of the control. A balanced value can 
be derived by evaluating how much one is ready to spend in energy to achieve a 
given performance. The ratio between performance gain and actuation cost of the 
most effective open-loop control suggests a value close to $\gamma$ = 3/2. The 
parameters used for GPC are summarized in Tab.~\ref{tab:pmmhparam}.

\begin{table}
{
\caption{GPC parameters used for the control of the PMMH backward facing step flow.}\label{tab:pmmhparam}
\begin{tabular}{cc}
\hline\hline
Parameter & Value\\\hline
n & 500\\
$P_r$ & 0.1\\
$P_m$ & 0.20\\
$P_c$ & 0.70\\
$n_p$ & 7\\
$n_e$ & 1\\
Node functions & $+,-,\times,/,\exp,\log,\tanh$\\
\hline\hline
\end{tabular}
}
\end{table}

\subsubsection{Results}
\begin{figure}
\includegraphics[width=.45\textwidth]{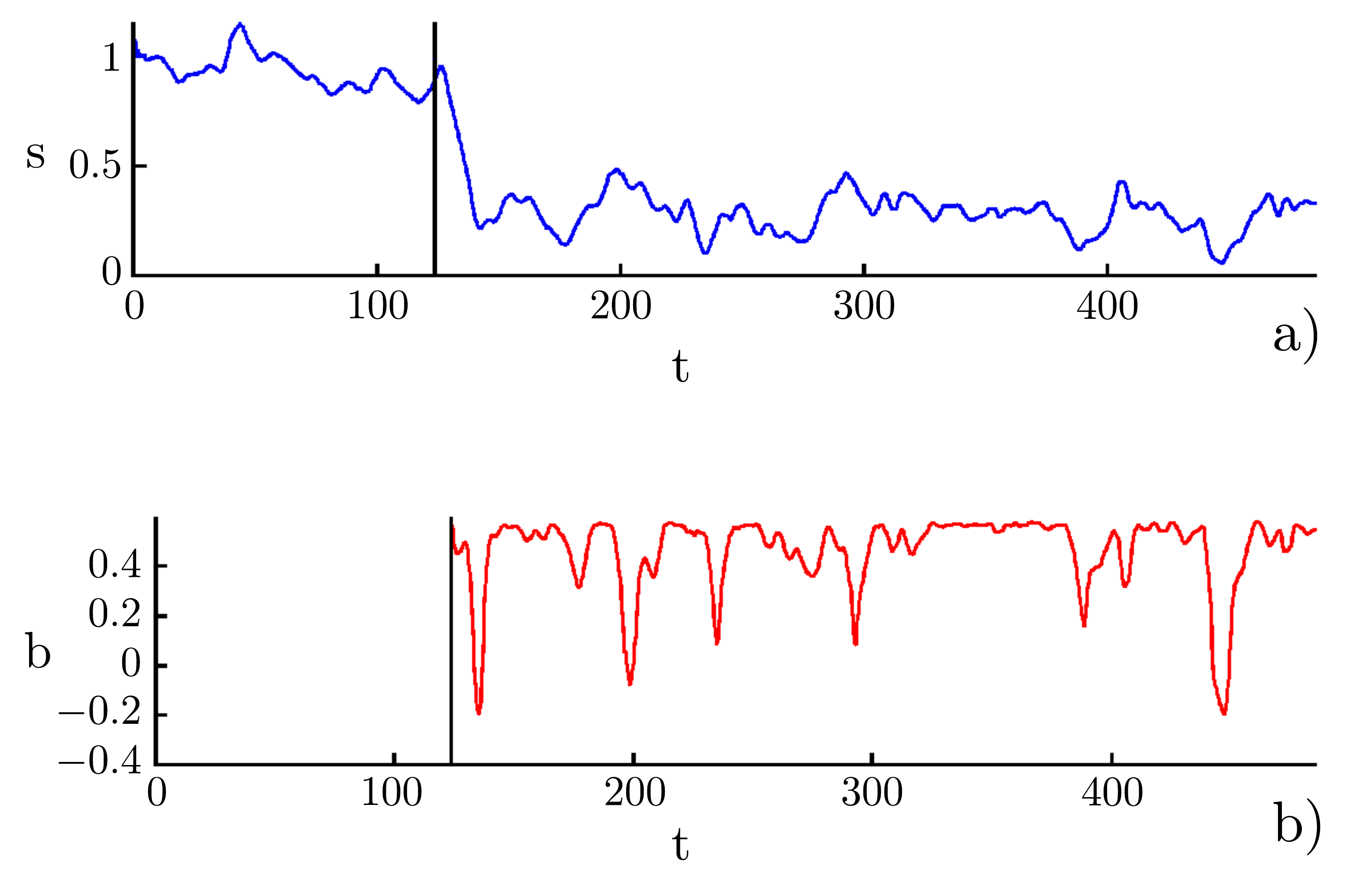}
\caption{Sensor signal $s$ for natural and GPC controlled flow (a) and control signal $b$ (b). The vertical black line shows when control starts. Time $t$ is non-dimentionnalized using step height and free-stream velocity.}\label{res:pmmh}
\end{figure}

The GPC process converges after 8 generations with 500 individuals each. Figure~\ref{res:pmmh} shows the time series of natural and GPC controlled flow. The law found by GPC ensures a $58\%$ reduction of the recirculation area, quite similarly to the best open-loop control at that Reynolds number. While the best open-loop case is an harmonic forcing around the vortex shedding frequency ($1\,\text{Hz})$, a frequency analysis of the control signal for the GPC case shows a dominant frequency around $0.1\,\text{Hz}$ indicating that two different physical processes are used for the same reduction. Once again, shifting the operating conditions shows that the closed-loop control determined by GPC is much more robust than any open-loop forcing (Tab.~\ref{tab:pmmh}): for both $Re_h=900$ and $1800$, GPC control manages to reduce the recirculation significantly (more than $50\%$ while the same open-loop forcing is inefficient with less than $10\%$ reduction out of the conditions where it has been designed). A detailed GPC study is described in~\cite{gautier2014jfm}.

\begin{table}
\caption{Performance of uncontrolled, best open-loop controlled and GPC controlled systems. While GPC performance is close to best open-loop control at the learning Reynolds number, the closed-loop control shines in maintaining performance in out of design conditions.}\label{tab:pmmh}
\begin{tabular}{c c c c}
\hline\hline
\multirow{2}{*}{case} & \multirow{2}{*}{natural} & best open-loop  & GPC \\
& & ($Re_h=1350$) & ($Re_h=1350$)\\
\hline
J ($Re_h=900$) & $1$ & $0.75$ & $0.33$\\
J ($Re_h=1350$) & $1$ & $0.42$ & $0.42$\\
J ($Re_h=1800$) & $1$ & $0.76$ & $0.59$\\
\hline\hline
\end{tabular}
\end{table}

\subsection{Wall turbulence control in LML experiment}\label{sec:LML}
GPC has been applied in the wall-turbulence wind tunnel at LML (Fig.~\ref{fig:exp_id}c) with the goal to reattach the flow after the natural separation point occurring after a sharp ramp. The performances of the GPC have been compared to those of the best open-loop scheme obtained from an
extensive parametric study.

\subsubsection{Experimental setup}
\begin{figure}
\centering
\includegraphics[width=0.4\textwidth]{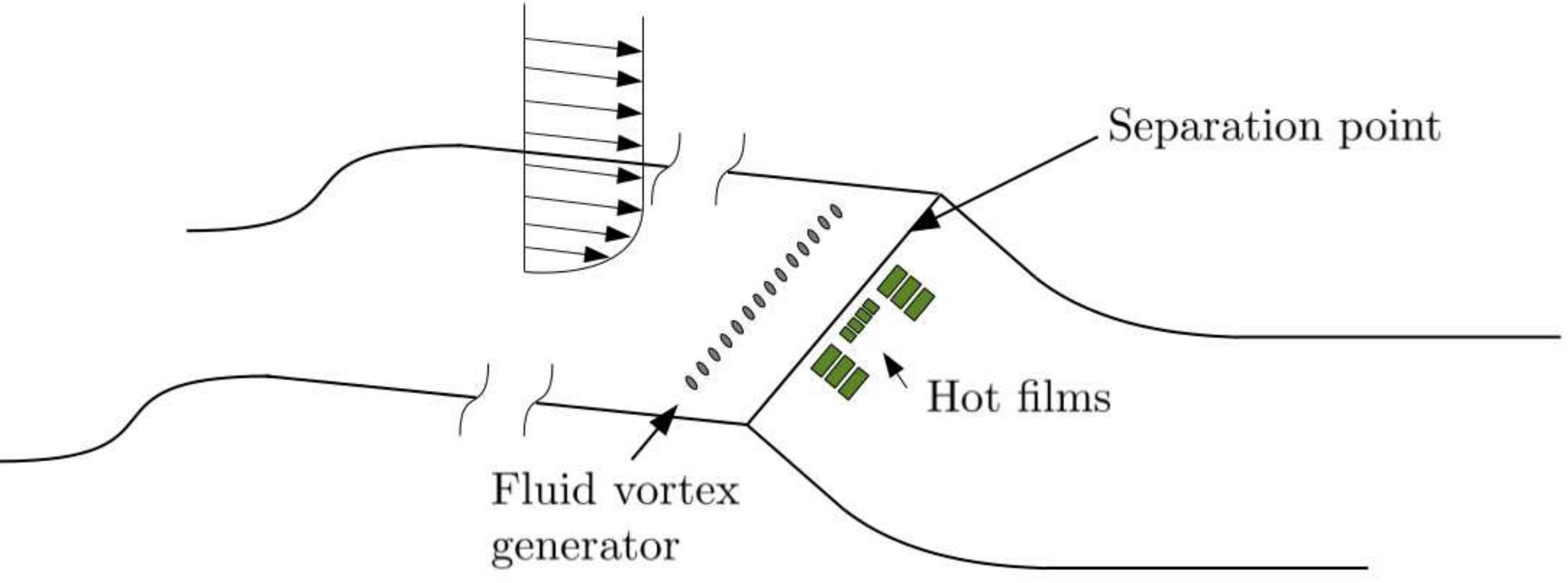}
\caption{Experimental configuration of the separating boundary layer in the LML wind tunnel. An array of co-rotating fluid vortex generators is placed upstream of the separation. The hot-film rake allows to qualify the friction at the wall, indicating potential re-attachment.}
\label{fig:LML_expsetup}
\end{figure}

The LML wall-turbulence wind tunnel is a $1\times 2 \,\text{m}^2$ cross-section wind tunnel operated with a flow velocity up to $10\,\text{m.s}^{-1}$. The AVERT ramp model~\cite{Cuvier2014JoT} is placed in the test section (Fig.~\ref{fig:LML_expsetup}b). This ramp was designed to select the pressure gradient of the upcoming boundary layer. A sharp edge is placed at the end of the ramp in order to force a separation of the flow at this fixed position. The actuation is made of $6\,\text{mm}$ diameter angled jets ($35^{\circ}$ pitch angle, $125^{\circ}$ skew angle) placed throughout the span, upstream of the separation in order to produce an optimal array of co-rotating streamwise vortices in the boundary layer~\cite{cuvier2012controle}. These can be triggered in with an on/off mode using electro-valves with a maximal frequency of $300\,\text{Hz}$.
Hot-films are placed in the descending part of the bump in order to record the wall friction at that position. These are used as sensors for the GPC process. Three hot-films are selected thanks to their sensitivity during previous open-loop and closed-loop attempts. We define the sensors signals as:
\begin{equation}
s_i=\frac{h_i-h_{i,u}}{h_{i,\text{max}}-h_{i,u}},\quad \text{with}\quad i=A,B,C, \label{eq:LMLsensor}
\end{equation}
where $h_i$ is the raw voltage output of sensor $i$, $h_{i,u}$ the average voltage for the uncontrolled case (corresponding to a separated flow and low friction) and $h_{i,\text{max}}$ the average voltage for the most efficient forcing, leading to maximal friction, in the case of constant blowing. As can be seen, the effectiveness of the actuation in reattaching the flow is here characterized by maximizing the wall friction (it was verified in previous experiments that the sign of the wall friction does not play a role~\cite{cuvier2012controle}).
The goal function used in this experiment is aimed at maximizing the friction while penalizing the actuation cost:
\begin{equation}
J=\left(\frac{1}{3}\sum_{i=A,B,C} \left<s_i\right>_T^2\right)^{-1} + \gamma \left<b\right>_T^2,\label{eq:JLML}
\end{equation}
with $b$ the actuation value (0 or 1) and $\gamma=2$ a penalization coefficient. Sensors A, B and C are hot-films placed the closest to the separation line at different span-wise position. Finally a PIV plane is setup in the symmetry plane to assess the effectiveness of the reattachment~\cite{Cuvier2014JoT}. The parameters used for GPC are summarized in Tab.~\ref{tab:lmlparam}.

\begin{table}
{
\caption{GPC parameters used for the control of the LML separating boundary layer}\label{tab:lmlparam}
\begin{tabular}{cc}
\hline\hline
Parameter & Value\\\hline
n & 500\\
$P_r$ & 0.1\\
$P_m$ & 0.25\\
$P_c$ & 0.65\\
$n_p$ & 7\\
$n_e$ & 1\\
Node functions & $+,-,\times,/,\exp,\log,\tanh$\\
\hline\hline
\end{tabular}
}
\end{table}

\subsubsection{Results}
\begin{figure}
\begin{center}
\includegraphics[width=0.4\textwidth]{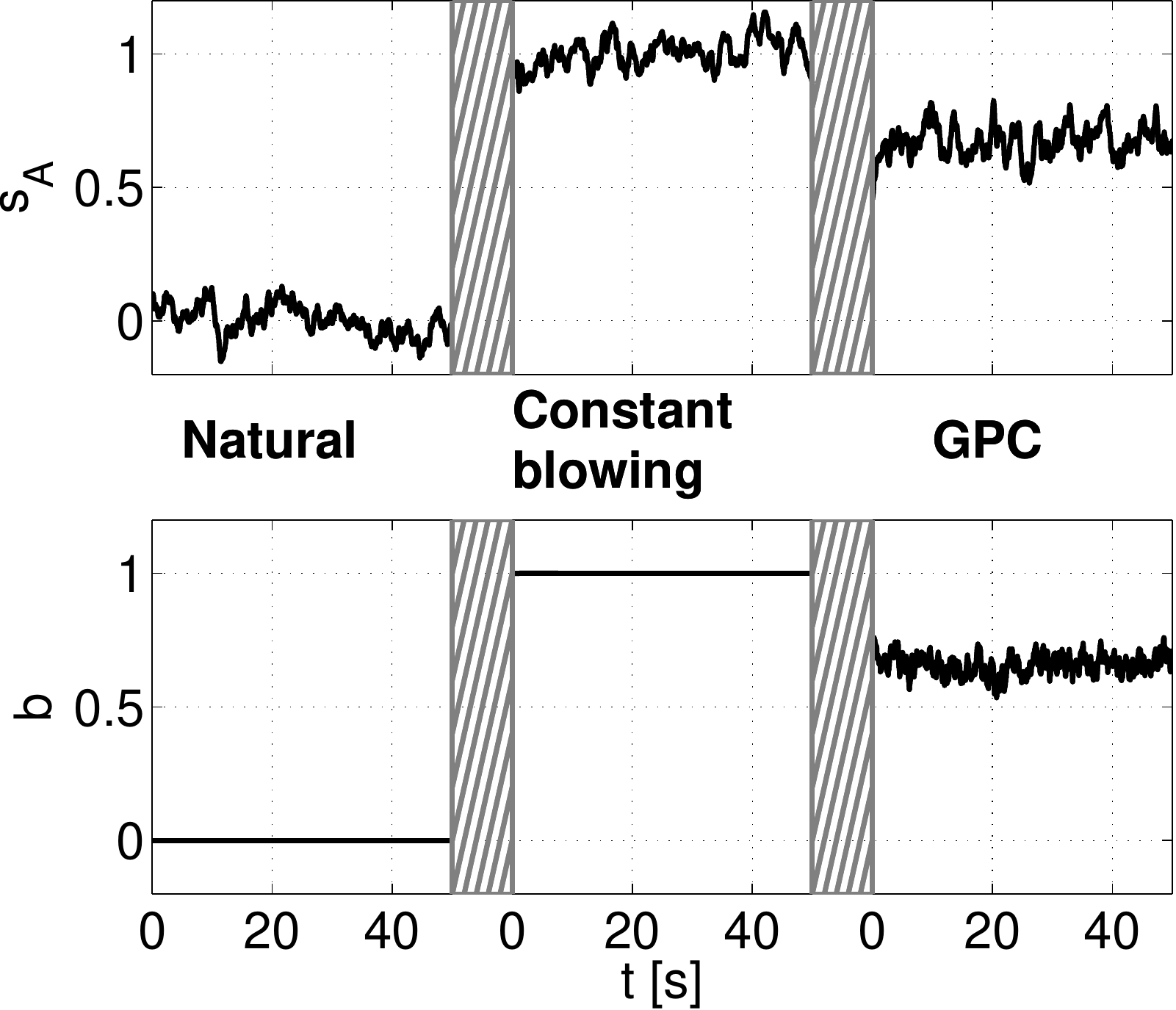}
\end{center}
\caption{Time series of sensor A (top) and low-pass filtered control signal (bottom) for uncontrolled (left), open-loop constant blowing control (center) and GPC control (right). }
\label{fig:lml}
\end{figure}

\begin{figure}
\begin{center}
\begin{tabular}{c c}
\includegraphics[width=0.4\textwidth]{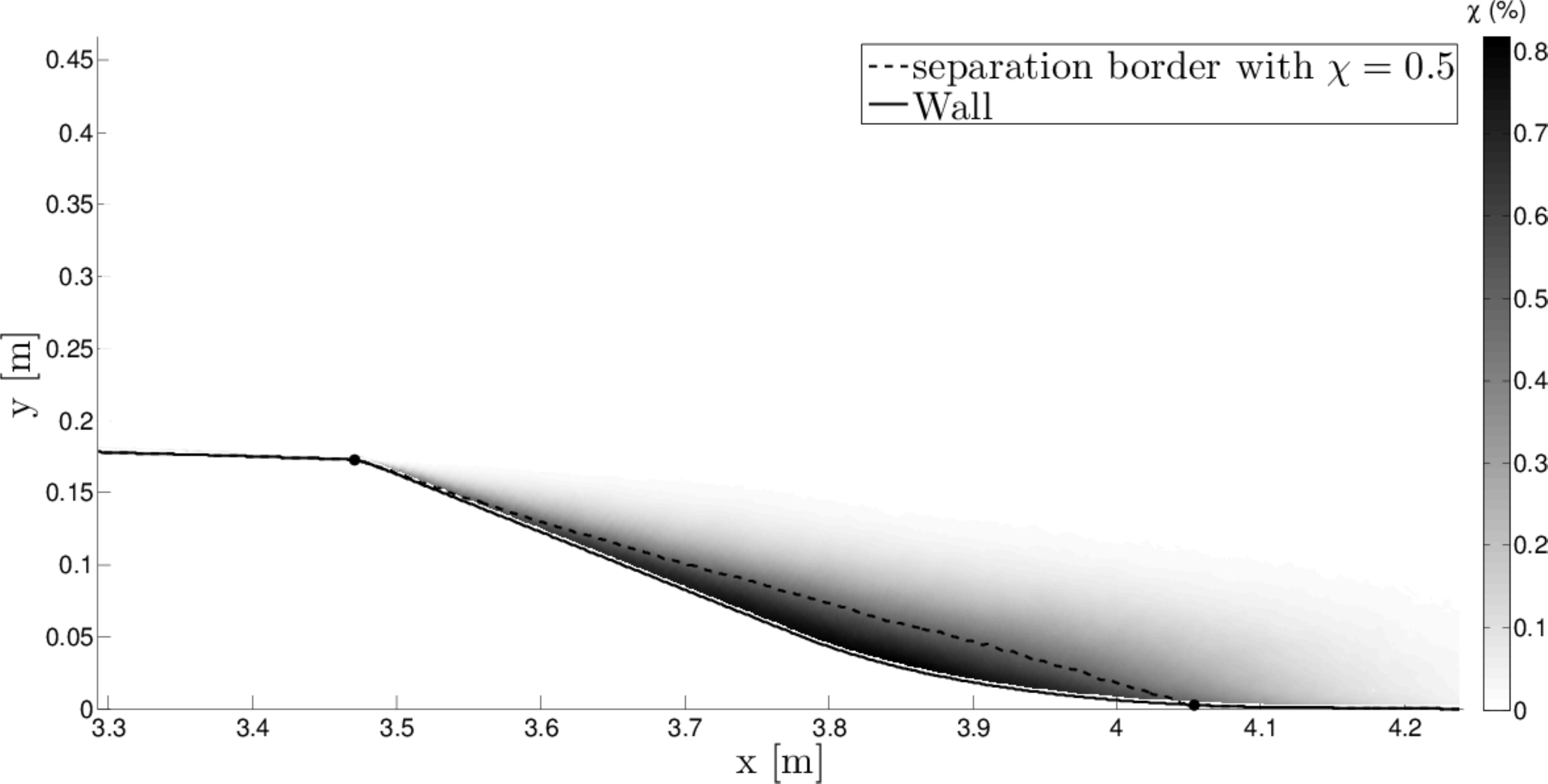}& a)\\
\includegraphics[width=0.4\textwidth]{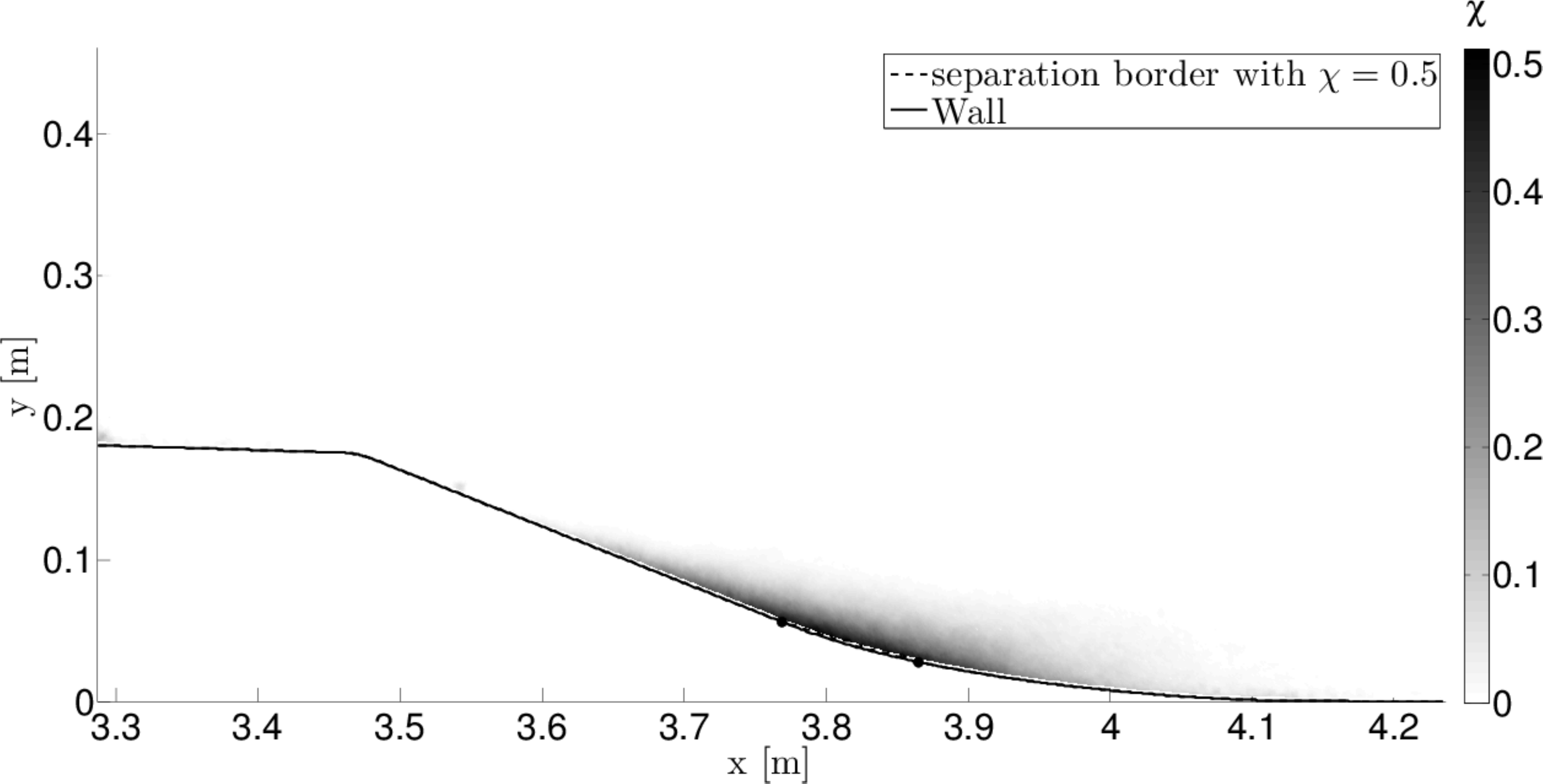}& b)
\end{tabular}
\end{center}
\caption{Cartographies of back-flow coefficient $\chi$, the ratio of the negative over positive value for the streamwise velocity at the considered point~\cite{simpson1989ARFM}. a: Uncontrolled flow. b: GPC controlled flow. The separation has been drastically reduced. For both cases the iso-line at $\chi=50\,\%$ has been traced.}
\label{fig:lmlpiv}
\end{figure}

Natural, constantly forced and GPC controlled time series of the A sensor and filtered actuation are displayed in Fig.~\ref{fig:lml}. According to \eqref{eq:JLML} and \eqref{eq:LMLsensor}, the $J$ value for the natural case is infinite as $\left<h_i\right>_T=h_{i,u}$ in this case. We observe a $33\%$ reduction of $J$ in the GPC case compared to the constant blowing actuation (Tab.~\ref{tab:lml}). Nevertheless Fig.~\ref{fig:lml} indicates that the friction is lower in the GPC case than in the constant blowing case (though the reattachment is effective, see Fig.~\ref{fig:lmlpiv}). All previous studies have shown that at the selected operating conditions, the influence of the frequency in the actuation is limited. The parametric study with respect to the duty cycle of an harmonic forcing reveals that above a certain frequency the relation between duty cycle and friction is monotonic with a positive slope. In few words, the more you blow,  the more you re-attach. Previous control attempts revealed that the constant blowing is actually the best control configuration with respect to wall friction~\cite{cuvier2012controle}. By penalizing the actuation strength in the control objective function, we actually select an operating point (which is reproducibly reached by several successive attempts) while applying the GPC process. All the control laws selected by GPC exhibit the same behavior: a rather high frequency content in the actuation signal with an average duty cycle around $70\%$. Different parameters and conditions have been tried and will be summed-up in an upcoming communication.

\begin{table}
\caption{Respective J values for uncontrolled, best open-loop constant blowing control and GPC control. The uncontrolled $J$ value is theoretically infinite and the constant blowing has its $J$ value fixed by construction at 3. GPC achieves a $33\%$ reduction of the J value.}\label{tab:lml}
\begin{tabular}{c c c c}
\hline\hline
case & natural & best open-loop & GPC\\
\hline
J & $\infty$ & $3$ & $2.1$\\
\hline\hline
\end{tabular}
\end{table}

\subsection{Wall turbulence control in PRISME experiment}\label{sec:PRISME}
GPC has been applied in the PRISME wall-bounded turbulence experiment with the goal to reattach the flow after the natural separation point occurring downstream a sharp ramp. A picture of the experimental setup is given in Fig.~\ref{fig:exp_id}d. The design of PRISME experiment is
close to that of the LML described in the previous subsection. Both experiments are involved in the SepaCoDe project funded
by the French National Research Agency (ANR). These experiments differ by the oncoming boundary layer relative to the ramp
height, the operating free-stream velocity, the physical time-scales and the properties of the unsteady vortex generators.
The performances of the GPC have been compared to those of the best open-loop scheme obtained from an
extensive parametric study.

\subsubsection{Experimental setup}
The experiments have been carried out in the Malavard closed-loop wind tunnel at the PRISME laboratory~\cite{Debienetal2014}. The square test-section is $2\,\text{m}$ wide and 5 m long.
A settling chamber and a strong contraction located upstream the test section ensure a residual turbulence intensity as low as $0.4\%$.
\begin{figure}[htbp]
\centering
\includegraphics[width=0.4\textwidth]{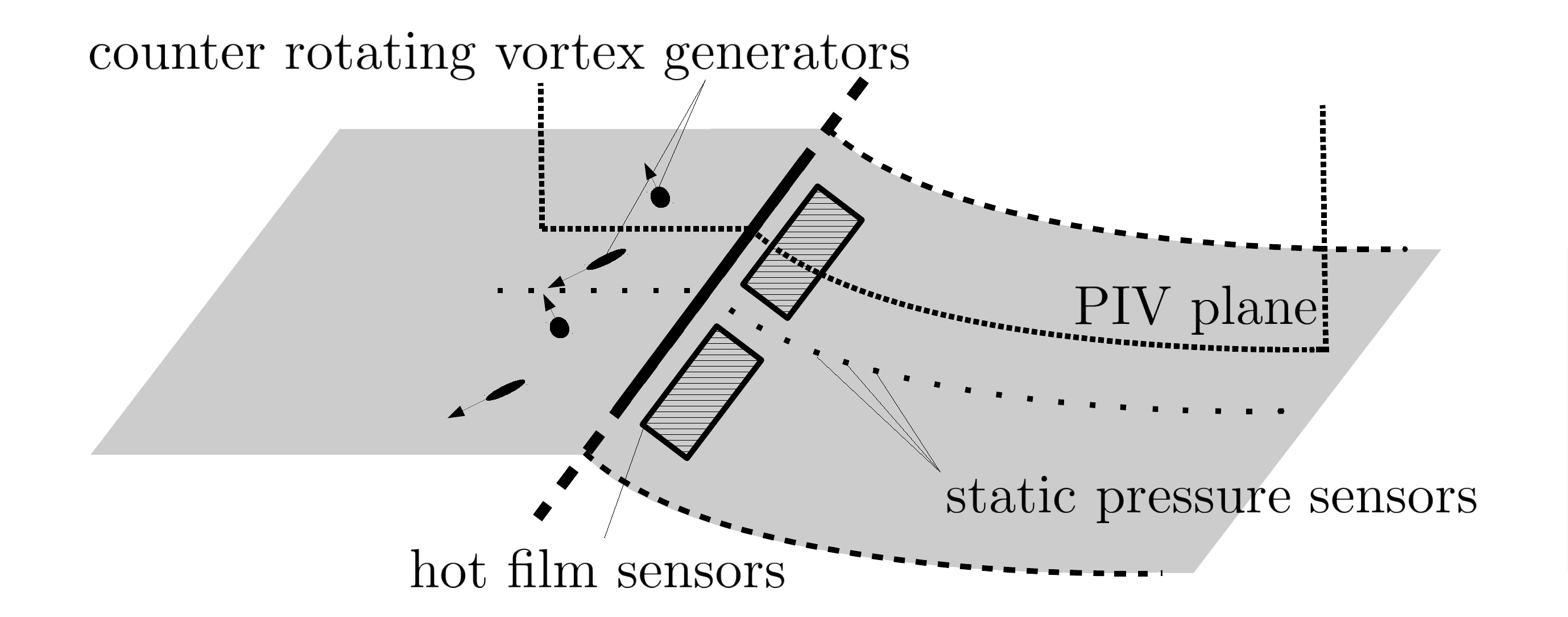}
\caption{Experimental configuration of the separating boundary layer in the PRISME wind tunnel. The jets are placed to generate counter-rotating streamwise vortices. Hot-film sensors are placed after the separation and static pressure sensors are located in the symmetry plane.\label{fig:UVG}}
\end{figure} 
A model is installed at the middle height of test section and it spans the entire width of the test section. A massive separation
is produced by a backward-facing ramp whose height and length are $h = 100$ mm and $\ell = 470$ mm, respectively. The onset of
the separation is fixed by a sharp edge with a 25\textdegree\ slant angle. The free-stream velocity $U_\infty$ is $20 \text{m.s}^{-1}$
yielding a ramp-height-based Reynolds number $Re_h = U_\infty h / \nu$ of around $1.3 \times 10^5$.  

For control purpose, $54$ counter-rotating unsteady vortex generators (UVG) have been implemented one boundary layer
thickness upstream of the sharp edge ramp (see Fig.~\ref{fig:UVG}). Their design and location have been extrapolated from the results
reported by \cite{GodardStanislas2006, Shaqarinetal2013, Cuvieretal2011}. The jet velocity ratio is $V_{jet}/U_{\infty} = 3$.
Each actuator is composed of two counter-rotating jets each with hole diameter of $\phi= 1.2$\,mm. They present a pitch angle of 
$\alpha = 135$\textdegree\, are skewed by $\beta = 45$\textdegree\ and have a distance of $\lambda/\phi = 15$
between them. The distance between the center line of two consecutive VGA is $L/\phi = 30$.

Similarly to section~\ref{sec:LML}, temporal variations of wall-shear-stress are recorded by means of two hot-films located
on the ramp together with the jet actuation collected from a mass flow controller. Furthermore, the pressure distribution along
the model is acquired in order to compute the global properties of the flow. A PIV equipment is also used in order to record the recirculation length (see Fig.~\ref{fig:UVG}).

The sensors $s_i$ used for the GPC control law are based on the hot-film signals:
\begin{equation}
s_i = \frac{h_i - h_{i,u}}{h_{i, {\rm max}} - h_{i,u}}, \mbox{with}\, i = A, B,
\end{equation}
where $h_i$ is the raw voltage output of sensor $i$, $h_{i,u}$ the average voltage for the uncontrolled case (corresponding to a separated flow and low friction) and $h_{i, {\rm max}}$ the average voltage for the most efficient forcing, leading to maximal friction, in the case of constant blowing.

The evaluation of the individual is achieved according to the following cost function:
\begin{equation}
J =  J_{\rm HF} + \gamma_{\rm p_{\rm stat}} J_{\rm p_{\rm stat}} + \gamma_{\rm act} J_{\rm act},
\label{eq:J_final}
\end{equation}
with $J_{\rm HF}$ being an evaluation of the friction recorded from the hot-films, $J_{\rm p_{\rm stat}}$ an evaluation based
on the static pressure distribution and $J_{\rm act}$ an evaluation of the actuation cost. $\gamma_{\rm p_{\rm stat}}=1/200$
and $\gamma_{\rm act}=0.6$ stand for penalization coefficients. The evaluation based on the friction is defined as:
\begin{equation}
J_{\rm HF} = \frac{1}{N_{\rm HF}} \sum_{i=1}^{N_{\rm HF}}\left[ 1-
              \tanh\left(\frac{\langle {\rm HF}_i\rangle_T}{\langle {\rm HF}_{i,0}\rangle_T}-1\right)\right],
  \label{eq:J_HF}
\end{equation}
where $N_{\rm HF}=2$ is the number of hot-film sensors, ${\rm HF}_i$ the sensor value collected from the $i^{th}$ hot-film and ${\rm HF}_0$ the
hot-film sensor value when no actuation is present. The value of $J_{\rm HF}$ is 1 when no effect is recorded and approaches 0 as the friction
increases. The evaluation based on the static pressure is defined as:
\begin{equation}
J_{\rm p_{\rm stat}} = \left<\frac{1}{0.1 + \sum_i (p(x_i)-p_u(x_i))^2 
             \frac{x_{\rm max}-x_i}{x_{\rm max}-x|_{x=0}} }\right>_T,
  \label{eq:J_p}
\end{equation}
with $x_i$ the position of the $i^{th}$ pressure tap after the edge, $x|_{x=0}$ the position of the pressure tap closest to the edge, $x_{\rm max}$ the furthest downstream pressure tap, $p(x_i)$ the static pressure recorded at position $x_i$ and $p_u(x_i)$ the static pressure
recorded at position $x_i$ in the uncontrolled case. $\left<\cdot\right>_T$ is the average over the evaluation time $T$. $J_{\rm p_{\rm stat}}$ is equal to 10 when both controlled and uncontrolled pressure
distributions collapse and goes to zero when they deviate from each other, with a linearly increasing weight when approaching the separation
point. The evaluation of the actuation cost is defined as:
\begin{equation}
J_{\rm act}=\left<\frac{Q}{Q_u}\right>_T,
\end{equation}
where $Q$ is the flow-rate and $Q_u$ the flow-rate under constant blowing. $J_{\rm act}$ is equal to 1 for constant
blowing and is null when no actuation is recorded. The parameters used for GPC are summarized in Tab.~\ref{tab:prismeparam}.

\begin{table}
{
\caption{GPC parameters used for the control of the PRISME separating boundary layer}\label{tab:prismeparam}
\begin{tabular}{cc}
\hline\hline
Parameter & Value\\\hline
n & 100\\
$P_r$ & 0.1\\
$P_m$ & 0.2\\
$P_c$ & 0.7\\
$n_p$ & 7\\
$n_e$ & 1\\
Node functions & $+,-,\times,/,\exp,\log,\tanh$\\
\hline\hline
\end{tabular}
}
\end{table}

\subsubsection{Results}  
\begin{figure}[htbp]
\centering
\subfigure[]
{\includegraphics[width=0.4\textwidth]{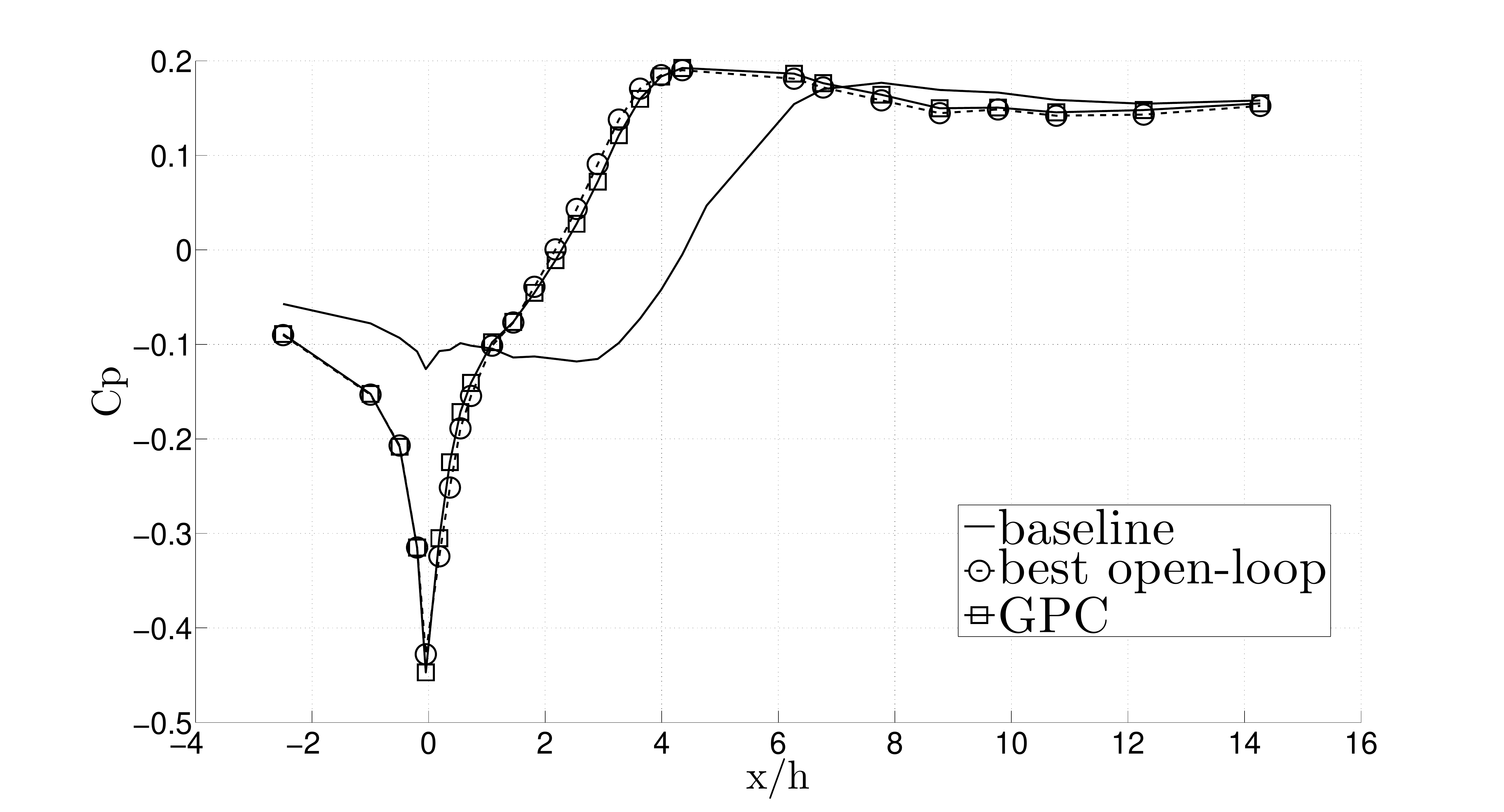}
\label{fig:Pstat}}
\hspace{0.2cm}
\subfigure[]
{\includegraphics[width=0.4\textwidth]{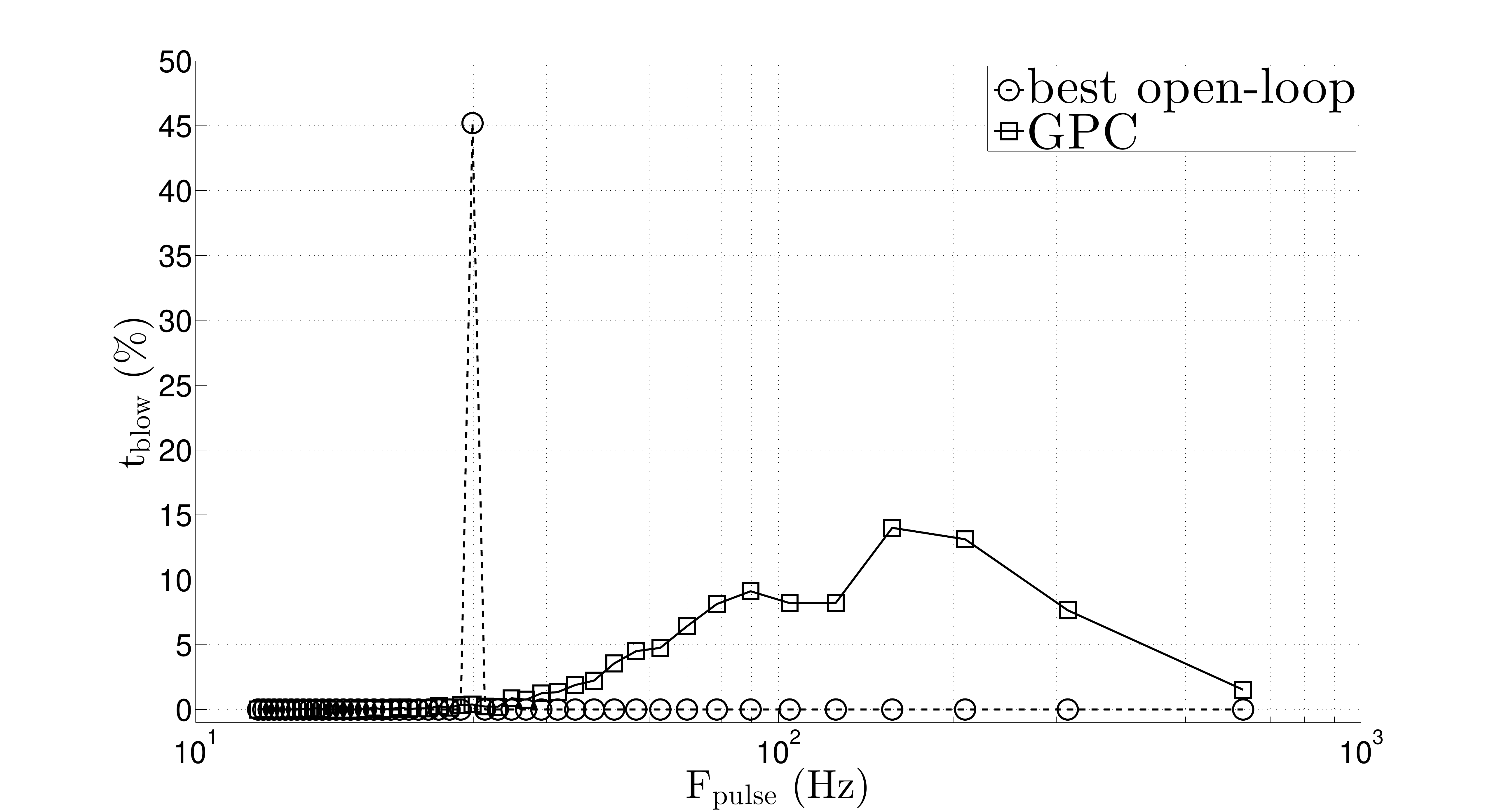}
\label{fig:Tblow}}
\caption{(a) Pressure distribution along the ramp. (b) Frequency distribution of the blowing.}
\end{figure}

The pressure distribution obtained in controlled cases is compared to that of the baseline flow in Fig.~\ref{fig:Pstat}. Both open-loop
and GPC schemes lead to a reduction of the mean recirculation region since the recovery region associated to the pressure plateau
is shifted upstream. Noticeable is the acceleration of the flow induced by the UVGs upstream the sharp edge location ($x/h=0$) as emphasized
by the strong decrease in pressure. However, pressure distributions computed for the best open-loop and GPC almost collapse meaning that from
a global viewpoint, the efficiency of both control approaches are mostly equivalent. This is confirmed by the measurement of the separation
length $L_{sep}$ from the PIV dataset which is reduced by about 40\% when control is applied (see Tab.~\ref{tab:Prop}). Nevertheless, the actuation
cost to achieve the same separation reduction is significantly lower ($\approx 20$\%) for the GPC as evidenced by the momentum coefficients $c_\mu$
($\sim \frac{S_j d_c V_j^2}{1/2 S_{\rm ref} U_\infty^2}$ with $S_j$ the total cross section of the jets, $d_c$ the duty cycle and $S_{\rm ref}$ the surface of
the ramp) reported in Tab.~\ref{tab:Prop}.
\begin{table}
\caption{Cost function values, separation and actuation properties for the natural, best open-loop and GPC cases.}
\begin{tabular}{cccc}
\hline\hline
Case & natural & open-loop & GPC \\\hline
J &    50.4    &      0.291        &	 0.32\\
$L_{sep}/h$ & 5.4 & 3.14 & 3.16 \\
$c_\mu$ ($\times 10^{-4}$) & - & 16.51 & 13.66\\
\hline\hline
\end{tabular}
\label{tab:Prop}       
\end{table}

One of the main differences of the actuation between our best open-loop case and the GPC relates on the frequency distribution (computed from a
zero-crossing algorithm applied on the mass flow controller signal) of the blowing (see Fig.~\ref{fig:Tblow}). Unlike open-loop control for which
the blowing frequency is unique (here 30 Hz), the frequency distribution of the GPC is broad-band. More surprisingly, the frequency range of the
GPC does not match that of the best open-loop case meaning that even though both control approaches lead to the same separation reduction, their
underlying mechanisms differs significantly. A more detailed study of the flow physics will be reported in an upcoming publication. 

\section{Conclusions}\label{sec:conclusion}
We propose a model-free optimization of sensor-based control laws 
for general multiple-input multiple-output (MIMO) plants, 
the 'Genetic Programming Control' (GPC).
This model-free strategy, based on genetic programming, is aimed at offering a solution for the in-time closed-loop control of complex non-linear systems like turbulent flows. 
GP is one of the most versatile methods for function optimization in machine learning.
GPC relies on the evolution of an ensemble (called 'generation')
of general nonlinear functions (called 'individuals')
and invests in exploring the function-space.
Since GPC is an evolutionary algorithm, it has a large chance to detect and exploit otherwise invisible local extrema.
In contrast, 
model-free adaptive control is particularly suited 
for adjusting one or few parameters 
of prescribed open- or closed-loop control laws to changing flow conditions.
Such on-line parameter adaptation
is not part of the presented GPC method 
but could --- in principle --- be included.

As our first test-case,
GPC has been successfully applied 
to a closed-loop stabilization of a oscillator model 
detecting and exploiting frequency cross-talk in an unsupervised manner. 
Frequency cross-talk is of crucial importance 
for large-scale turbulence control
with complex interactions between the coherent structures at different dominant frequencies: 
the mean flow changes on  large time scales while the cascade to small-scale structures has small associated time scales.
By definition, frequency cross-talk is ignored in any linearized system. Another successful demonstration of GPC 
is closed-loop control for the maximization of
the Lyapunov exponent (stretching) 
of the forced Lorenz equations.
Again, this increase of unpredictability is a highly nonlinear phenomenon.

Most importantly, GPC has also been applied successfully to a panel of experiments representative of the challenges encountered in modern turbulent fluid dynamics. In all cases GPC has been able to derive an effective control law in a model-free approach and with performances consistently beating the optimized comparison control studies. All of these control laws are closed-loop control laws which exhibit an inherent robustness when compared with open-loop forcing which is by far the most commonly used strategy for the control of turbulent flows. Furthermore GPC has sometimes derived control laws that exploit an unexpected mechanism as in the PMMH experiment on the backward facing step. GPC has proved to be effective in exploring the search-space characterized for each control problem by its cost function. 

While it can be claimed that having to evaluate all individuals from each generation is a major time investment, the total time investment is not larger than with other methods, and is in fact mostly lower. First of all the approach is model free, which means that no prior experiments are needed, neither to perform a system identification or corroborate the model adequacy to the plant. GPC can be applied on any closed-loop ready system with very few preparations consisting mainly in implementing a function that can translate the individual and creating the communications to exchange the individuals and their cost values. With the accumulated experience, this is now a few hours task. Then, in most unfavorable case, i.e. in the PMMH water tunnel where evaluation time has to be larger due to the small velocities when compared to air flows, and where 500 individuals per generation have been employed, a full week of experiment has been employed. This means that only a week was needed to derive an effective in time closed-loop control law for an experimental flow. To our knowledge this beats any other approach. In wind tunnels, with the velocities involved in the considered studies (of the order of $10\,\text{m.s}^{-1}$) and while dealing with small generations (50 individuals), a control law can be obtained within around 8 hours of operation. 

While the optimality of the results from evolutionary algorithms can never be proved, the goal of the method never was to obtain the best possible control law, but obtaining one which is performing well within a range of the cost function values decided by the operator. GPC shines when no modeling of the flow can be obtained in finite time, thus obtaining a performing control law is already a major success.

GPC has overcome important technical challenges for in-time control:
(1) the sensors show broadband frequency dynamics,
(2) there is a large convective time delay from actuators to sensors and 
(3) the responses were found to be strongly nonlinear.

Summarizing, the model-free formulation of GPC gives rise to a high flexibility: 
it can be applied to any MIMO plant and use any cost function.
Though a model is not needed, 
the more we know about the system, 
the better we can design the cost function according to the underlying physics
and the better we can bias the control law selection.
Further improvements can be expected from including actuation or sensor histories,
like in ARMAX models \citep{Herve2012jfm}.
The relation of tree depth, number of generations, number of individuals with convergence is subject of ongoing research and may boost the performance considerably. 

The model-free control design is particularly interesting 
for experimental applications for which a model might not even be known,
like for the control of some multi-phase or multi-physics flows 
with several phases, combustion or unknown non-Newtonian fluids. The following step in machine learning control will certainly be to extract knowledge from the results of GPC in order to build models for the flows by studying the set of best control laws obtained. This way the model-free GPC will play a major role in determining non-linear models for complex flows.
We conjecture that GPC will play a similar role as control theory
in the closed-loop control of turbulence and other complex flows.

\begin{acknowledgments}
We acknowledge funding of the French Science Foundation ANR on projects Chaire 
d'Excellence TUCOROM (ANR-10-CHEX-0015) and SepaCoDe (ANR-11-BS09-0018). MS and 
MA acknowledge the support of the LINC project (no. 289447) funded by EC’s 
Marie-Curie ITN program (FP7-PEOPLE-2011-ITN).
We thank Eurika Kaiser and Nathan Kutz for fruitful discussions and comments.
\end{acknowledgments}


%

\end{document}